\documentclass{article}
\usepackage{amsmath,amsfonts,amsthm,color,amssymb,mathtools}
\usepackage{graphicx}
\usepackage{hyperref}

\newtheorem{theorem}{Theorem}
\theoremstyle{remark}
\newtheorem{remark}{Remark}
\usepackage[normalem]{ulem}

\usepackage{geometry}
\usepackage{enumerate}

\usepackage{tikz}
\usepackage{float}
\usetikzlibrary{matrix,shapes,chains,positioning,decorations.pathreplacing,arrows}

\newcommand{\R}{\mathrm{R}}

\newcommand{\ind}{1\hspace{-2.1mm}{1}} 
\newtheorem{definition}{Definition}

\DeclareMathOperator*{\argmin}{\arg\!\min}
\begin{document}
\title{Deep Learning Volatility\\
\Large{A deep neural network perspective on pricing and calibration in (rough) volatility models}\thanks{The authors are grateful to Jim Gatheral, Ben Wood, Antoine Savine and Ryan McCrickerd for stimulating discussions. MT conducted research within the « Econophysique et Syst\`emes Complexes » under the aegis of the Fondation du Risque, a joint initiative by the Fondation de l'\'Ecole Polytechnique, l'\'Ecole Polytechnique, Capital Fund Management. MT also gratefully acknowledges the financial support of the ERC
679836 Staqamof and the Chair Analytics and Models for Regulation.}}
\author{Blanka Horvath\\
\large{Department of Mathematics, King's College London}
\\
\normalsize{blanka.horvath@kcl.ac.uk, bhorvath@turing.ac.uk}
 \and Aitor Muguruza \\
\large{Department of Mathematics, Imperial College London
 \& NATIXIS}
\\
\normalsize{aitor.muguruza-gonzalez15@imperial.ac.uk}\\ 
\and Mehdi Tomas\\
\large{CMAP \& LadHyx, \'Ecole Polytechnique}\\
\normalsize{mehdi.tomas@polytechnique.edu}
}
\maketitle
\begin{abstract}
\noindent We present a neural network based calibration method that performs the calibration task within a few milliseconds for the full implied volatility surface. The framework is consistently applicable throughout a range of volatility models---including the rough volatility family---and a range of derivative contracts.
The aim of neural networks in this work is an off-line approximation of complex pricing functions, which are difficult to represent or time-consuming to evaluate by other means. We highlight how this perspective opens new horizons for quantitative modelling: The calibration bottleneck posed by a slow pricing of derivative contracts is lifted. This brings several numerical pricers and model families (such as rough volatility models) within the scope of applicability in industry practice. The form in which information from available data is extracted and stored influences network performance.
This approach is inspired by representing the implied volatility and option prices as a collection of pixels. In a number of applications we demonstrate the prowess of this modelling approach regarding accuracy, speed, robustness and generality and also its potentials towards model recognition.
\\

\noindent \textbf{2010 }\textit{Mathematics Subject Classification}: 60G15, 60G22, 91G20, 91G60, 91B25
\end{abstract}

\tableofcontents
\newpage
\section{Introduction}

Approximation methods for option prices came in all shapes and forms in the past decades and they have been extensively studied in the literature and well-understood by risk managers. Clearly, the applicability of any given option pricing method (Fourier pricing, PDE methods, asymptotic methods, Monte Carlo, \ldots etc.) depends on the regularity properties of the particular stochastic model at hand. Therefore, tractability of stochastic models has been one of the most decisive qualities in determining their popularity. In fact it is often a more important quality than the modelling accuracy itself: It was the (almost instantaneous) SABR asymptotic formula that helped SABR become the benchmark model in fixed income desks, and similarly the convenience of Fourier pricing is largely responsible for the popularity of the Heston model, despite the well-known hiccups of these models. Needless to say that it is the very same reason  (the concise Black Scholes formula) that still makes the Black-Scholes model attractive for calculations even after many generations of more realistic and more accurate stochastic market models have been developed. On the other end of the spectrum are rough volatility models, for which (despite a plethora of modelling advantages, see \cite{BFG15,ER18, GJR14} to name a few) the necessity to rely on relatively slow Monte Carlo based pricing methods creates a major bottleneck in calibration, which has proven to be a main limiting factor with respect to industrial applications.
This dichotomy can become a headache in situations when we have to weigh up the objectives of accurate pricing vs. fast calibration against one another in the choice of our pricing model. 
In this work we explore the possibilities provided by the availability of an algorithm that --for a choice of model parameters-- directly outputs the corresponding vanilla option prices (as the Black-Scholes formula does) for a large range of maturities and strikes of a given model.\\

\noindent In fact, the idea of mapping model parameters to shapes of the implied volatility surface directly is not new. The stochastic volatility inspired SSVI, eSSVI surfaces (see \cite{Gatheral04,GJ14,HM17}) do just that: A given set of parameters is translated directly to different shapes of (arbitrage-free) implied volatility surfaces, bypassing the step of specifying any stochastic dynamics for the underlying asset. For stochastic models that admit asymptotic expansions, such direct mappings from model parameters to (approximations of) the implied volatility surface in certain asymptotic regimes can be obtained (one example is the famous SABR formula). Such asymptotic formulae are typically limited to certain asymptotic regimes along the surface by their very nature.
Complementary to asymptotic expansions we explore here a direct (approximative) mapping from different parameter combinations of stochastic models to different shapes of implied volatility surface for intermediate regimes. It's appeal is that it combines the advantages of direct parametric volatility surfaces (of the SSVI family) with the possibility to link volatility surfaces to the stochastic dynamics of the underlying asset.\\

In this paper we apply deep neural networks (merely) as powerful high-dimensional functional approximators to approximate the multidimensional pricing functionals from model parameters to option prices. The advantage of doing so via deep neural networks  over standard (fixed-basis) functional approximations is that deep neural networks are agnostic to the approximation basis \cite{MLBook}. This makes them robustly applicable to several stochastic models consistently. Our objective in doing so is to move the (often time-consuming) numerical approximation of the pricing functional into an off-line preprocessing step. This preprocessing amounts to storing the approximative direct pricing functional in form of the network weights after a supervised training procedure:
Using available numerical approximations of option prices as ground truth (in a stochastic model of our choice), we train a neural network to learn an accurate approximation of the pricing functional. After training, the network outputs--for any choice of model parameters--the corresponding implied volatilities within milliseconds for a large range of maturities and strikes along the whole surface. 
Furthermore, we show that this procedure generalises well for unseen parameter combinations: the accuracy of price approximation of our neural network pricing functional on out-of-sample data is within the same range as the accuracy of the original numerical approximation used for training.
The accuracy of this direct pricing map is demonstrated in our numerical experiments.\\

One of the striking advantages of this approach is that it speeds up the (on-line) calibration Rough Volatility models to the realm of just a few milliseconds. There have been several recent contributions on neural network calibrations of stochastic models \cite{BS18, BLTW2018,Hernandez, KON2018, MadanSchoutens}. Clearly, much depends on the finesse of the particular network design with respect to the performance of these networks. 
One contribution of this paper is to achieve a fast and accurate calibration of the rough Bergomi model of \cite{BFG15} with a general forward variance curve (approximated by piecewise constant function). To demonstrate this, we first perform calibration experiments on simulated data and show calibration accuracy in controlled experiments. To demonstrate the speed and prowess of the approach we then calibrate the rough Bergomi model to historical data and display the evolution of parameters on a dataset consisting of 10 years of SPX data.\\
Another advantage of our modelling choice is that by its very design it can be applied to portfolios including multiple strikes and maturities at the same time which is the first step towards their application as hedging instruments. See for example Buehler et al. \cite{BLTW2018} a motivation.\\

\noindent The paper is organised as follows:
In Section \ref{sec:Newperspectivecalib} we present a neural network perspective on model calibration and recall stochastic models that are considered in later sections. In this section we also formalise our objectives about the accuracy and speed of neural network approximation of pricing functionals and the basic ideas of our training. 
Section \ref{sec:neuralnetworks} recalls some background on neural networks as functional approximators and some aspects of neural network training that influenced the setup of our network architecture and training design. 
In Section \ref{sec:architecture} we describe the network architecture and training of the price approximation network such as the calibration methods we consider.
In Section \ref{sec:Numerics} we present numerical experiments of price approximations of vanilla and some exotic options, calibration to synthetic data and to historical data. Section \ref{sec:Conclusions} points to further potential applications and outlook to future work.\\
Numerical experiments and codes are provided on \href{https://github.com/amuguruza/NN-StochVol-Calibrations}{GitHub: NN-StochVol-Calibrations }, where an accessible code demo of our results can be downloaded. We also created a library of stochastic models where this approach is demonstrated to work well.

\newpage
\section{A neural network perspective on model calibration}\label{sec:Newperspectivecalib}

In plain words, any calibration procedure is meant to fix the model parameters such that the model is as close as possible to the observed reality. In a financial context, our model represents the underlying (stocks, indices, volatility, etc.) and we are interested in calibrating the model to available market prices of financial contracts based on this underlying.\\\
\ Let us first formalise this by setting the notation
$\mathcal{M}:=~\mathcal{M}(\theta)_{\theta\in\Theta}$ which represents an abstract model with parameters $\theta$ in the set $\Theta \subset \R^{n}$, for some $n\in \mathbb{N}$.  Thus the model $\mathcal{M}(\theta)$ (stochastic or parametric) and the corresponding prices of financial contracts are fully specified by the choice of the parameter combination $\theta \in \Theta$.
Furthermore, we introduce a pricing map $P:\mathcal{M}(\theta,\zeta)\to R^m$, where $\zeta:\left(C(\mathbb{R})\rightarrow \R^{m}\right)$, $m \in \mathbb{N}$ denote the financial products we aim to price, such as vanilla options for (a set of) given maturities and strikes. Let us denote the observed market data corresponding to the contracts, by $\mathcal{P}^{MKT}(\zeta)\in\mathbb{R}^{m}$, \ $m \in \mathbb{N}$.\\

\noindent \textbf{Parameter Calibration}:
The parameter configuration $\hat{\theta}$ solves an (ideal) $\delta$-calibration problem for a model $\mathcal{M}(\Theta)$ for the conditions $\mathcal{P}^{MKT}(\zeta)$ if 
\begin{equation}\hat{\theta}=\argmin_{\theta\in\Theta}\delta(P(\mathcal{M}(\theta),\zeta),\mathcal{P}^{MKT}(\zeta))\label{eq:CalibrationProblem}
\end{equation}
where $\delta(\cdot,\cdot)$ is a suitable choice of metric for the financial contract $\zeta$ at hand.

\noindent For most financial models however \eqref{eq:CalibrationProblem} represents an idealised form of the calibration problem as in practice there rarely exists an analytical formula for the option price $P(\mathcal{M}(\theta),\zeta)$ and for the vast majority of financial models it needs to be computed by some numerical approximation scheme.\\

\noindent \textbf{Approximate Parameter Calibration}
We say that the parameter configuration $\hat{\theta} \in \Theta$ solves an \textit{approximate $\delta$-calibration} problem for the model $\mathcal{M}(\Theta)$ for the conditions $\mathcal{P}^{MKT}(\zeta)$ if 
\begin{equation}\hat{\theta}=\argmin_{\theta\in\Theta} \delta(\widetilde{P}(\mathcal{M}(\theta),\zeta),\mathcal{P}^{MKT}(\zeta))\label{eq:ApproxCalibrationProblem}
\end{equation}
where $\delta(\cdot,\cdot)$ is a suitably chosen metric and $\widetilde{P}$ is a numerical approximation of the pricing map $P$.\\

\noindent In the remainder of this paper it is this second type of calibration problem that we will be concerned with: In our numerical experiments (Section \ref{sec:Numerics}) we consider the numerical approximation $\widetilde{P}$ of the pricing map $P$ as the benchmark (available truth) for generating synthetic training samples in the training a neural network to approximate pricing maps. Clearly, the better the original numerical approximations, the better the network approximation will be. In a separate work we will illuminate this perspective with a Bayesian analysis of the calibration procedure.
\subsection{A brief reminder of some (rough) models considered}
\label{sec:stochastic models}
\label{sec:pricingI}
We would like to emphasize that our methodology can in principle be applied to any (classical or rough) volatility model. From the classical Black Scholes or Heston models to the rough Bergomi model of \cite{BFG15}, also to large class of rough volatility models (see Horvath, Jacquier and Muguruza \cite{HJM17} for a general setup). In fact the methodology is not limited to stochastic models, also parametric models of implied volatility could be used for generating training samples of abstract models, but we have not pursued this direction further. \\\\
\textbf{The Rough Bergomi model}\\
In the abstract model framework, the rough Bergomi model is represented by $\mathcal{M}^{rBergomi}(\Theta^{rBergomi})$, with parameters $\theta = (\xi_0,\nu,\rho,H)\in \Theta^{rBergomi}$. On a given filtered probability space $(\Omega, \mathcal{F}, (\mathcal{F}_t)_{t\geq0}, \mathbb{P})$ the model corresponds to the following system
\begin{align}\label{example:rBergomi}
\begin{split}
dX_t&=-\frac{1}{2} V_t dt +\sqrt{V_t} dW_t,\quad \textrm{for} \ t>0, \quad X_0=0, \\
V_t&=\xi_0(t)\mathcal{E}\left(\sqrt{2H}\nu \int_0^t (t-s)^{H-1/2}dZ_s\right),\quad \textrm{for} \ t>0, \quad V_0=v_0>0
\end{split}
\end{align}
where $H\in (0,1)$ denotes the Hurst parameter, $\nu>0$ , $\mathcal{E}(\cdot)$ the stochastic exponential \cite{Doleans}, and $\xi_0(\cdot) >0$ denotes the initial forward variance curve (see \cite[Section 6]{BergomiBook}), 
and $W$ and $Z$ are correlated standard Brownian motions with correlation parameter $\rho\in [-1,1]$. To fit the model parameters into our abstract model framework $\Theta^{rBergomi}\subset \mathbb{R}^n$ for some $n \in \mathbb{N}$, the initial forward variance curve $\xi_0(\cdot)>0$ is approximated  by a piecewise constant function in our numerical experiments in Sections \ref{sec:rBergomiGeneral}, and  \ref{sec:rBergomiGeneralCalib}. We refer the reader to Horvath, Jacquier and Muguruza \cite{HJM17} for one general setting of rough volatility models and their numerical simulation.\\

\noindent \textbf{The Heston model}\\
The Heston model, appearing in our numerical experiments of Section \ref{sec:Conclusions} is described by the system
\begin{align}\label{example:Heston}
\begin{split}
dS_t&=\sqrt{V_t}S_tdW_t\quad \textrm{for} \ t>0, \quad S_0=s_0\\
dV_t&=a(b-V_t)dt+v \sqrt{V_t}dZ_t \quad \textrm{for} \ t>0, \quad V_0=v_0
\end{split}
\end{align}
with $W$ and $Z$ Brownian motions with correlation parameter $\rho\in [-1,1]$, $a,b,v>0$ and $2ab>v^2$. In our framework it is denoted by $\mathcal{M}^{Heston}(\theta)$ with $\theta=(a,b,v,\rho) \in \Theta^{Heston} \subset \mathbb{R}^4$. The Heston model is considered in our numerical experiments in Section \ref{sec:Conclusions}. It was also considered by \cite{BS18, HestonConvolutional} in different neural network contexts.\\

\noindent \textbf{The Bergomi model}\\
In the general $n$-factor Bergomi model, the  volatility is expressed as
\begin{align}\label{example:nFBergomi}
\begin{split}
V_t= \xi_0(t) \mathcal{E}\left(\eta_i \sum_{i=1}^n \int_0^t \exp\left(-\kappa_i (t-s) dW_s^i\right) \right)\quad \textrm{for} \ t>0, \quad V_0=v_0>0,
\end{split}
\end{align}
where $\eta_1, \ldots, \eta_n >0$ and $(W^1,\ldots,W^n)$ is an $n$-dimensional correlated Brownian motion, $\mathcal{E}(\cdot)$ the stochastic exponential \cite{Doleans}, and $\xi_0(\cdot) >0$ denotes the initial forward variance curve, see \cite[Section 6]{BergomiBook} for details. In this work we consider the Bergomi model for $n=1,2$ in Section \ref{sec:Numerics}.
Henceforth, $\mathcal{M}^{1FBergomi}(\xi_0,\beta,\eta,\rho)$ represents the 1 Factor Bergomi model, corresponding to the following dynamics:
\begin{align}\label{example:1FBergomi}
\begin{split}
dX_t&=-\frac{1}{2} V_t dt +\sqrt{V_t} dW_t\quad \textrm{for} \ t>0, \ X_0=0 \\
V_t&=\xi_0(t)\mathcal{E}\left(\eta \int_0^t \exp(-\beta(t-s))dZ_s\right)\quad \textrm{for} \ t>0, \quad V_0=v_0>0,
\end{split}
\end{align}
where $\nu >0$, 
and $W$ and $Z$ are correlated standard Brownian motions with correlation parameter $\rho\in [-1,1]$. To fit the model parameters into our abstract model framework $\Theta^{1FBergomi}\subset \mathbb{R}^n$, for some $n\in \mathbb{N}$, the initial forward variance curve $\xi_0(\cdot)>0$ is approximated in our numerical experiments by a piecewise constant function in Sections \ref{sec:rBergomiGeneral}, and  \ref{sec:rBergomiGeneralCalib}.\\

\noindent \textbf{The SABR model}\\
The stochastic alpha beta rho model of Hagan et al. \cite{Hagan1, Hagan2} is denoted in our setting as $\mathcal{M}^{SABR}(\alpha,\beta,\rho)$ and is defined as
\begin{align}\label{example:SABR}
\begin{split}
dS_t&=V_tS_t^{\beta}dW_t\quad \textrm{for}\ t>0, \quad S_0=s_0.\\
dV_t&=\alpha V_tdZ_t\quad \textrm{for} \ t>0, \quad V_0=v_0
\end{split}
\end{align}
where $v_0,s_0,\alpha >0$ and $\beta\in[0,1]$.
The SABR model is considered by McGhee in \cite{McGhee} in a neural network context (see also Section \ref{BypassingBottleneck}).

\subsection{Calibration bottlenecks in volatility modelling and deep calibration}

\noindent Whenever for a stochastic volatility model the numerical approximate calibration procedures \eqref{eq:ApproxCalibrationProblem} are computationally slow, a bottleneck in calibration time can deem the model of limited applicability for industrial production irrespective of other desirable features the model might have. This is the case in particular for the family rough volatility models, where the rough fractional Brownian motion in the volatility dynamics rules out usual Markovian pricing methods such as finite differences. 
So far such calibration bottlenecks have been a major limiting factor for the class of rough volatility models, whose overwhelming modelling advantages have been explored and highlighted in rapidly expanding number of academic articles \cite{AGM18, AlosLeon,BFG15,BFGMS17, BFGHS, BLP16, ER16, Fukasawa, GJR14, JMM17, HJL,  JPS17} in the past years.
Other examples include models with delicate degeneracies (such as the SABR model around zero forward) which for a precise computation of arbitrage-free prices require time consuming numerical pricing methods such as Finite Element Methods \cite{HR18}, Monte Carlo \cite{OsterleeI,OsterleeII} or the evaluation of multiple integrals \cite{Antonov}. \\ 


\noindent Contrary to Hernandez's \cite{Hernandez} pioneering work, where he develops a direct calibation via NN, we set up and advocate a two setp calibration approach.\\\\
\noindent\textbf{Two Step Approach (i) Learn a model and (ii) Calibrate to data:} One separates the calibration procedure described in \eqref{eq:ApproxCalibrationProblem} (resp. \eqref{eq:ApproxCalibrationProblem}) into two parts: \textbf{(i)} We first learn (approximate) the pricing map by a neural network that maps parameters of a stochastic model to pricing functions (or implied volatilities (cf. section (\ref{sec:pricingI}) and we store this map during an off-line training procedure. In a second step \textbf{(ii)} we calibrate (on-line) the now deterministic approximative learned price map, which speeds up the on-line calibration by orders of magnitude. To formalise the two step approach, we write for a payoff $\zeta$ and a model $\mathcal{M}$ with parameters $\theta\in \Theta$
\begin{align}\label{eq:SeparatedCalibrationProblem}
\textbf{(i) Learn: } \widetilde{F}(\Theta, \zeta)
=\widetilde{P}(\mathcal{M}(\Theta, \zeta))\qquad  \textbf{(ii) Calibrate: } \hat{\theta}=\argmin_{\theta\in\Theta} \delta(\widetilde{F}(\theta, \zeta),\mathcal{P}^{MKT}(\zeta)).
\end{align}
Note that in part \textbf{(ii)} of \eqref{eq:SeparatedCalibrationProblem} we essentially replaced $\widetilde{P}(\mathcal{M}(\Theta, \zeta))$ in equation \eqref{eq:ApproxCalibrationProblem} by its learned (deterministic) counterpart  $\widetilde{F}(\Theta, \zeta)$ (which will be a Neural Network see Section \ref{sec:architecture}) from \textbf{(i)}.
Therefore, this second calibration is--by its deterministic nature--considerably faster than calibration of all those traditional stochastic models, which involve numerical simulation of the expected payoff ${P}(\mathcal{M}(\theta, \zeta))=\mathbb{E}[\zeta(X({\theta}))]$ for some underlying stochastic process $X^{\theta}$.
The first part \textbf{(i)} in \eqref{eq:SeparatedCalibrationProblem} denotes an approximation of the pricing map through a neural network, which is calibrated in a supervised training procedure using the original (possibly slow) numerical pricing maps for training (see sections \ref{sec:architecture} and \ref{sec:Numerics} for details in specific examples). \\\\
\noindent In the following sections we elaborate on the objectives and advantages of this two step calibration approach and present examples of neural network architectures, precise numerical recipes and training procedures to apply the two step calibration approach to a family of stochastic volatility models. We also present some numerical experiments (corresponding codes are available on \href{https://github.com/amuguruza/NN-StochVol-Calibrations}{GitHub: NN-StochVol-Calibrations }) and report on learning errors and on calibration times.

\subsection{Challenges in neural network approximations of pricing functionals}\label{BypassingBottleneck}
In general problem \eqref{eq:CalibrationProblem} and henceforth \eqref{eq:ApproxCalibrationProblem} is solved using suitable numerical optimisation techniques such as gradient descent \cite{MLBook}, specific methods for certain metrics (such as Lavenberg-Marquadnt \cite{Levenberg} for $L^2$), neural networks, or tailor-made methods to the complexity of the optimisation problem and objective function at hand\footnote{For details and an overview on calibration methods see \cite{MLBook}.}. But irrespective of their level of sophistication all optimisers for calibration share a common property: repeated (iterative) evaluation of the pricing map $\theta \mapsto P(\mathcal{M}(\theta), \zeta)$ (resp. an approximation $\widetilde{P}$ thereof) on each instance $\theta$ of consecutive parameter combinations until a sufficiently small distance $\delta(\widetilde{P}(\mathcal{M}(\theta),\zeta),\mathcal{P}^{MKT}(\zeta)$ 
between model prices and observed prices is achieved.
 Consequently, the pricing map is arguably the computational cornerstone of a calibration algorithm. Main differences between specific calibration algorithms effectively lie in the way the specific choice of evaluated parameter combinations $\{\theta_1,\theta_2\ldots\}$ are determined, which hence determines the total number $N$ of functional evaluations of the pricing function $\big(P(\mathcal{M}(\theta_i), \zeta)\big)_{i =1 \ldots N}$ used in the calibration until the desired precision $\delta(\widetilde{P}(\mathcal{M}(\hat{\theta}),\zeta),\mathcal{P}^{MKT}(\zeta))$ is achieved.
In case the pricing map 
\begin{align*}
P(\mathcal{M}(\cdot),\zeta) \ :\ &\Theta \longrightarrow P(\mathcal{M})\\
&\theta \mapsto P(\mathcal{M}(\theta), \zeta)
\end{align*}
involved in \eqref{eq:CalibrationProblem} is available in closed form, and can be evaluated instantaneously, the calibration \eqref{eq:ApproxCalibrationProblem} is fast even if a high number $N$ of functional evaluations is used.
If the pricing map is approximated numerically, calibration time depends strongly on the time needed to generate a functional evaluation of the numerical approximation 
\begin{align}\label{eq:numericalapprox}
\theta_i \mapsto \widetilde{P}(\mathcal{M}(\theta_i),\zeta), \qquad\theta_i \in \{\theta_1, \ldots \theta_N\}
\end{align} at each iteration $i=1,\ldots,N$ of the calibration procedure. Slow functional evaluations potentially cause substantial bottlenecks in calibration time.
This is where we see the most powerful use of the prowess of neural network approximation:\\

\noindent A neural network is constructed to replace in \textbf{(i)} of \eqref{eq:SeparatedCalibrationProblem} the pricing map, that is to approximate (for a given financial contract $\zeta$) the pricing map from the full set\footnote{Note that the set $\theta_1,\ldots, \theta_N$ in \eqref{eq:numericalapprox} is extended to the full set of possible parameter combinations $\Theta$ in \eqref{eq:NetworkPriceApproximator}.} of model parameters $\Theta$ of the model to the corresponding prices $P(\mathcal{M}(\theta, \zeta))$.  
The \emph{first challenge} for the neural network approximator of pricing functionals is to speed up this process and enable us to obtain \emph{faster functional evaluations} and thereby lift the bottleneck of calibration.
The \emph{second challenge}  
is to do so with an accuracy that remains within the error bounds of the original numerical pricing discretisation:\\
\begin{align}\label{eq:NetworkPriceApproximator}
\begin{split}
\widetilde{F}:\ &\Theta \longrightarrow \widetilde{P}(\mathcal{M})\\
&\theta \mapsto \widetilde{F}(\theta, \zeta)
\end{split}
\end{align}
More precisely (motivated by \eqref{eq:ApproxCalibrationProblem}), for any parameter combination $\theta \in \Theta$ we aim to approximate the numerical approximation $\widetilde{P}$ of the true option price $P$ with the neural network $\widetilde{F}$ up to the same order of precision $\epsilon>0$  up to which $\widetilde{P}$ approximates $P$. That is, for any $\theta \in \Theta$
\begin{align*}
\begin{split}
&\widetilde{F}(\theta)
=
P(\mathcal{M}(\theta),\zeta)+\mathcal{O}(\epsilon)\quad \text{whenever} \quad
\widetilde{P}(\mathcal{M}(\theta),\zeta)
=
P(\mathcal{M}(\theta),\zeta)+\mathcal{O}(\epsilon).
\\ 
\end{split}
\end{align*}
Therefore, our training objective is
\begin{align}\label{eq:ApproximatePricing}& \widetilde{F}(\theta)
=
\widetilde{P}(\mathcal{M}(\theta),\zeta)+\mathcal{O}(\epsilon).
\end{align}
where $\widetilde{P}$ is the available numerical approximation of the pricing function, which is considered as ground truth.
In our numerical experiments in Section \ref{sec:Numerics} we demonstrate that our approximation network achieves this approximation accuracy and yields a substantial speedup in terms of functional evaluations.

\subsection{Motivations for our choice of training setup and features of neural networks as approximators of pricing functionals}\label{sec:AdvantagesNN}
There are several advantages of separating the tasks of pricing and calibration  which we address in full detail in a separate work. Here we recall some of the most convincing reasons to do so. Above all, the most appealing reason is that it allows us to build upon the knowledge we have gained about the models in the past decades, which is of crucial importance from a risk management perspective. By its very design, deep learning the \emph{price approximation} \textbf{(i)} combined with \textbf{(ii)} deterministic calibration does not cause more headache to risk managers and regulators than the corresponding stochastic models do.
Designing the training as described above demonstrates how deep learning techniques can successfully extend the toolbox of financial engineering, without making compromises on any of our objectives.

\begin{enumerate}
\item The knowledge gathered in many years of experience with traditional models remains useful and risk management libraries of models remain valid. The neural network is only used as a computational enhancement of models. 

\item The availability of training data for training the deep neural network does not cause any constraints as it is synthetically generated by traditional numerical methods. 

\item This can be extended beyond the models presented in this work: Whenever a consistent numerical pricer exists for a model, it can be approximated and replaced by a deep neural network that provides fast numerical evaluations of the pricing map.
\end{enumerate}
Here, we identify the grid-based apporach as our choice of training. Though a thorough analysis of the best training approaches is subject to further research, we have good reason to believe that the grid-based approach provides a powerful and robust methodology for training:\\\\

\subsubsection{Reasons for the choice of grid-based implicit training}

In the grid-based approach we evaluate the values of implied volatility surface along $8\times 11$ gridpoints with $80,000$ different parameter combinations we effectively evaluate the "fit" of the surface to numerically generated ones across the same number of points. By moving the evaluation of the implied volatilities into the objective function we improve the learning in many aspects:
\begin{itemize}
\item The first advantage of implicit training is that it efficiently exploits the structure of the data. Updates in neighbouring volatility points $\sigma_{n-1}$ and $\sigma_n$ can be incorporated in the learning process. If the output is a full grid as in \eqref{eq:implicitNetwork} this effect is further enhanced. Updates of the network on each gridpoint also imply additional information for updates of the network on neighbouring gridpoints.  One can say that we regard the implied volatility surface as an image with a given number of pixels.
\item A further advantage of the image based implicit training is, that by evaluating the objective function on a larger set of (grid) points, injectivity of the mapping can be more easily guaranteed than in the pointwise training: Two distinct parameter combinations are less likely to yield the same value across a set of gridpoints, then if evaluated only on a single point. 
\item We do not limit ourselves to one specific grid on the implied volatility surface. We store the generated $60,000$ sample paths for the training data and chose a set of maturities (here 8) and strikes (here 11) to evaluate prices corresponding to these paths. But we can easily add and evaluate additional maturities and strikes to the same set of paths. Note in particular that in this training design we can refine the grid on the implied volatility surface without increasing the number of training samples needed and without significantly increasing the computational time for training as the portfolio of vanilla options on the same underlying grows with different strikes and maturities.
\end{itemize}

\subsubsection{Some relevant properties of deep neural networks as functional approximators}
\label{sec:neuralnetworks}
Deep feed forward\footnote{The network is called feed forward if there are no feedback connections in which outputs of the model are fed back into itself.} neural networks are the most basic deep neural networks, originally designed to approximate some function $F^{*}$, which is not available in closed form but only through sample pairs of given input data $x$ and output data $y=F^*(x)$. In a nutshell, a feed forward network defines a mapping $y=F(x,w)$ and the training determines (calibrates) the optimal values of network parameters $\widehat{w}$ that result in the best function approximation\footnote{
In our case $y$ is a $8\times 11$-point grid on the implied volatility surface and $x$ are model parameters $\theta \in \Theta$, for details see Section \ref{sec:pricing}.} $F^*(\cdot)\approx F(\cdot,\widehat{w})$ of the unknown function $F^*(\cdot)$ for the given pairs of input and output data $(x,y)$, cf.  \cite[Chapter 6]{MLBook}.
\\
To formalise this, we introduce some notation and recall some basic definitions and principles of function approximation via (feedforward) neural networks:
\begin{definition}[Neural network]
	Let $L \in \mathbb{N}$ and the tuple $(N_{1}, N_{2} \cdots, N_{L}) \in \mathbb{N}^{L}$ denote the number of layers (depth) and the number of nodes (neurons) on each layer respectively. 
Furthermore, we introduce the affine functions 
\begin{align}\label{eq:NNAffineFunction}\begin{split}
w^{l}: \ \mathbb{R}^{N_l}&\longrightarrow\mathbb{R}^{N_{l+1}} \textrm{ for } 1 \leq l \leq L-1\\
x&\mapsto A^{l+1}x+b^{l+1}
\end{split}\end{align} acting between layers for some $A^{l+1} \in \mathbb{R}^{ N_{l+1} \times N_l}$. The vector $b^{l+1} \in \mathbb{R}^{N_{l+1}}$ denotes the \emph{bias term} and each entry $A^{l+1}_{(i,j)}$ denotes the \emph{weight} connecting node $i \in N_l$ of layer $l$ with node $j \in N_{l+1}$ of layer $l+1$. For the the collection of affine functions of the form \eqref{eq:NNAffineFunction} on each layer we fix the notation $w = (w^1,\ldots,w^L)$. We call the tuple $w$ the \emph{network weights} for any such collection of affine functions.
Then a Neural Network $F(w,\cdot):\mathbb{R}^{N_0}\to \mathbb{R}^{N_L}$ is defined as the composition:
\begin{align}\label{eq:NeuralNetwork} 
F := F_{L} \circ \cdots \circ F_{1}
\end{align}
where each component is of the form $F_{l} := \sigma_{l} \circ W^{l}$.
The function $\sigma_{l}:\mathbb{R}\to\mathbb{R}$ is referred to as  the \textit{activation function}. It is typically nonlinear and applied component wise on the outputs of the affine function $W^{l}$.
The first and last layers, $F_{1}$ and $F_{L}$, are the \textit{input} and \textit{output} layers. Layers in between, $F_{2} \cdots F_{L-1}$, are called \textit{hidden layers}.
\end{definition}
\noindent The following central result of Hornik justifies the use of neural networks as approximators for multivariate functions and their derivatives.
\begin{theorem}[Universal approximation theorem (Hornik, Stinchcombe and White \cite{UniversalApprox})]\label{thm:UniveralApproximation}
Let $\mathcal{NN}^{\sigma}_{d_{0}, d_{1}}$ be the set of neural networks with activation function $\sigma: \mathbb{R} \mapsto \mathbb{R}$, input dimension $d_{0}\in\mathbb{N}$ and output dimension $d_{1}\in\mathbb{N}$. Then, if $\sigma$ is continuous and non-constant, $\mathcal{NN}^{\sigma}_{d_{0}, 1}$ is dense in $L^{p}(\mu)$ for all finite measures $\mu$.
\end{theorem}
\noindent There is a rapidly growing literature on approximation results with neural networks, see \cite{Hor91,UniversalApproxDerivatives,Mhas93,ShaClCo18} and the references therein. Among these we would like to single out one particular result:

\begin{theorem}[Universal approximation theorem for derivatives (Hornik, Stinchcombe and White \cite{UniversalApproxDerivatives})]\label{thm:UniveralApproximationDerivatives}
Let $F^*\in\mathcal{C}^n$ and $F: \mathbb{R}^{d_0}\to \mathbb{R}$ and $\mathcal{NN}^{\sigma}_{d_{0}, 1}$ be the set of single-layer neural networks with activation function $\sigma: \mathbb{R} \mapsto \mathbb{R}$, input dimension $d_{0}\in\mathbb{N}$ and output dimension $1$. Then, if the (non-constant) activation function is $\sigma\in\mathcal{C}^n(\mathbb{R})$, then $\mathcal{NN}^{\sigma}_{d_{0}, 1}$ arbitrarily approximates $f$ and all its derivatives up to order $n$.
\end{theorem}
\begin{remark}
Theorem \ref{thm:UniveralApproximationDerivatives} highlights that the smoothness properties of the activation function are of significant importance in the approximation of derivatives of the target function $F^*$. In particular, to guarantee the convergence of $l$-th order derivatives of the target function, we choose an activation function $\sigma \in C^l(\R)$. Note that the ReLu activation function , $\sigma_{ReLu}(x)=(x)^+$ is not in $\mathcal{C}^l(\R)$ for any $l>0$, while $\sigma_{Elu}(x)=\alpha (e^x-1)$ is smooth. 
\end{remark}

\begin{figure}[H]\label{im:singleneuron}
\begin{tikzpicture}[
init/.style={
  draw,
  circle,
  inner sep=2pt,
  font=\Huge,
  join = by -latex
},
squa/.style={
  draw,
  inner sep=4pt,
  font=\normalsize,
  join = by -latex
},
start chain=2,node distance=13mm
]
\node[on chain=2] 
  (x2) {$x_2$ 
  };
\node[on chain=2,join=by o-latex] 
  {$a_{i,2}$};
\node[on chain=2,init] (sigma) 
  {\normalsize$\displaystyle b_i+\sum_{j=1}^3 a_{i,j}x_j$};
\node[on chain=2,squa,label=above:{\parbox{2cm}{\centering Activation \\ function}}]   
  { $\sigma_{ELU}$};
\node[on chain=2,label=above:Output,join=by -latex] 
  {$y$};
\begin{scope}[start chain=1]
\node[on chain=1] at (0,1.5cm) 
  (x1) {$x_1$};
\node[on chain=1,label=above:Weights,join=by o-latex] 
  (w1) {$a_{i,1}$};
\end{scope}
\begin{scope}[start chain=3]
\node[on chain=3] at (0,-1.5cm) 
  (x3) {$x_3$};
\node[on chain=3,join=by o-latex] 
  (w3) {$a_{i,3}$};
\end{scope}
\node[label=above:\parbox{2cm}{\centering Bias \\ $b$}] at (4.65cm,2cm) (b) {};

\draw[-latex] (w1) -- (sigma);
\draw[-latex] (w3) -- (sigma);
\draw[o-latex] (b) -- (sigma);

\draw[decorate,decoration={brace,mirror}] (x1.north west) -- node[left=10pt] {Node inputs} (x3.south west);
\end{tikzpicture}
\caption{In detail neuron behaviour}
\end{figure}
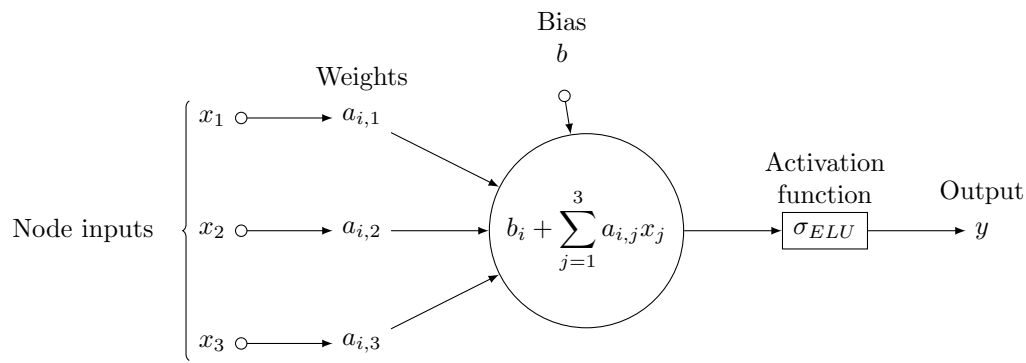

The following Theorem provides theoretical bounds for the above rule of thumb and establishes a connection between the number of nodes in a network and the number of training samples needed to train it.
\begin{theorem}[Estimation bounds for Neural Networks (Barron \cite{Barron})]
Let $\mathcal{NN}^{\sigma}_{d_{0}, d_{1}}$ be the set of single-layer neural networks with Sigmoid activation function $\sigma(x)=\frac{e^x}{e^x+1}$, input dimension $d_{0}\in\mathbb{N}$ and output dimension $d_{1}\in\mathbb{N}$. Then:
$$\mathbb{E}\|F^*-\hat{F}\|^2_2\leq\mathcal{O}\left(\frac{C_f^2}{n}\right)+\mathcal{O}\left(\frac{nd_0}{N}\log N\right)$$
where $n$ is the number of nodes, $N$ is the training set size and $C_{F^*}$ is the first absolut moment of the Fourier magnitude distribution of $F^*$.
\end{theorem}
\begin{remark}
Barron's \cite{Barron} insightful result gives a rather explicit decomposition of the error in terms of bias (model complexity) and variance:
\begin{itemize}
\item $\mathcal{O}\left(\frac{C_{F^*}^2}{n}\right)$ represents the model complexity, i.e. the larger $n$ (number of nodes) the smaller the error
\item $\mathcal{O}\left(\frac{nd_0}{N}\log N\right)$ represents the variance, i.e. a large $n$ must be compensated with a large training set $N$ in order to avoid overfitting.
\end{itemize}
\end{remark}

\noindent Finally, we motivate the use of multi layer networks and the choice of network depth. Even though a single layer might theoretically suffice to arbitrarily approximate any continuous function,in practice the use of multiple layers dramatically improves the approximation capacities of network. We informally recall the following Theorem due to Eldan and Shamir \cite{DepthNN} and refer the reader to the original paper for details.
\begin{theorem}[Power of depth of Neural Networks (Eldan and Shamir \cite{DepthNN})] \label{thm:depth}
There exists a simple (approximately radial) function on $\mathbb{ R}^d$, expressible by a small 3-layer
feedforward neural networks, which cannot be approximated by any 2-layer network, to more than a
certain constant accuracy, unless its width is exponential in the dimension.
\end{theorem}
\begin{remark}
In spite of the specific framework by Eldan and Shamir \cite{DepthNN} being restrictive, it provides a theoretical justification to the power of ``deep" neural networks (multiple layers) against ``shallower'' networks (i.e. few layers) as in \cite{McGhee} with a larger number of neurons. On the other hand, multiple findings indicate \cite{BS18,IS15} that adding hidden layers beyond 4 hidden layers does not significantly improve network performance.
\end{remark}

\section{Pricing and calibration with neural networks: Optimising network and training}\label{sec:pricing}
In this section we compare different objective functions (direct calibration to data to an image-based implicit learning approach) and motivate our choice of image-based objective function. We give details about network architectures for the approximation network and compare different optimisers for the calibration step.

\subsection{The objective function}
\begin{enumerate}
\item  Learn the map $F^*(\theta)=\{P^{\mathcal{M}(\theta)}(\zeta_i)\}_{i=1}^{n}$ via neral network, where $\{\zeta_i\}_{i=1,...,n}$ represents the exotic product attributes (such as maturity, strike or barrier level) on a prespecified grid with size $n$. 
$$\hat{w}=\displaystyle\argmin_{w\in\mathbb{R}^n}\sum_{u=1}^{N_{Train}}\sum_{i=1}^{n}  \left(F(\theta_u,w)_{i}-F^*(\theta_u)_{i}\right)^2.$$
\item Solve \begin{equation}\label{eq:generalOF}\hat{\theta}:=\displaystyle\argmin_{\theta\in\Theta}\sum_{i=1}^n (\widetilde{F}(\theta)_{i}-P^{MKT}(\zeta_i))^2.\end{equation}
\end{enumerate}
\subsubsection{For vanillas}
\noindent As in many academic and industry research papers, we pursue the calibration of vanilla contracts via approximation of the implied volatility surface\footnote{For sake of completeness we introduce the Black-Scholes Call pricing function in terms of log-strike $k$, initial spot $S_0$, maturity $T$ and volatility $\sigma$:
$$BS(\sigma,S_0,k,T):=S_0\mathcal{N}(d_+)-K\mathcal{N}(d_-),\quad d_{\pm}:=\frac{\log(S_0)-k}{\sqrt{T}\sigma}\pm \frac{\sqrt{T}\sigma}{2},$$
where $\mathcal{N}(\cdot)$ denotes the Gaussian cumulative distribution function. The implied volatility induced by a Call option pricing function $P(K,T)$ is then given by the unique solution $\sigma_{BS}(k,T)$ of the following equation
$$BS(\sigma_{BS}(k,T),S_0,k,T)=P(k,T).$$
Precisely, we seek to solve the following calibration problem
\begin{equation}\hat{\theta}:=\argmin_{\theta\in\Theta} d(\Sigma^{\mathcal{M}(\theta)}_{BS},\Sigma^{MKT}_{BS})\label{eq:ImpliedVolCalibration}\end{equation}
where $\Sigma^{\mathcal{M}(\theta)}_{BS}:=\{\sigma^{\mathcal{M}(\theta)}_{BS}(k_i,T_j)\}_{{i=1,..,n,\;j=1,...,m}}$ represents the set of implied volatilities generated by the model pricing function $P(\mathcal{M}(\theta),k,T)$ and $\Sigma^{MKT}_{BS}:=\{\sigma^{MKT}_{BS}(k_i,T_j)\}_{{i=1,..,n,\;j=1,...,m}}$ are the corresponding market implied volatilities, for some metric $d:\mathbb{R}^{n\times m}\times \mathbb{R}^{n\times m}\to \mathbb{R}^+ $.}. \\\\
We take this idea further and design an implicit form of the pricing map that is based on storing the implied volatility surface as an image given by a grid of "pixels". This image-based representation has a formative contribution in the performance of the network we present in Section \ref{sec:Numerics}. We present our contribution here; Let us denote  by $\Delta:=\{k_i,T_j\}_{i=1,\;j=1}^{n,\;\;\;\;m}$ a fixed grid of strikes and maturities, then we propose the following two step approach:
\begin{enumerate}
\item Learn the map $F^*(\theta)=\{\sigma^{\mathcal{M}(\theta)}_{BS}(T_i,k_j)\}_{i=1,\;j=1}^{n,\;\;\;\;m}$ via neural network $\widetilde{F}(\theta):=F(\theta,\hat{w})$ where
\begin{align}F^*:\Theta&\longrightarrow \mathbb{R}^{n\times m}\label{eq:implicitNetwork}\\ \theta &\mapsto F^*(\theta) \nonumber\end{align}
where the input is a parameter combination $\theta\in\Theta$ of the stochastic model $\mathcal{M}(\Theta)$ and the output is a $n\times m$ grid on the implied volatility surface $\{\sigma^{\mathcal{M}(\theta)}_{BS}(T_i,k_j)\}_{i=1,\;j=1}^{n,\;\;\;\;m}$ where $n,m\in\mathbb{N}$ are chosen appropriately (see Section \ref{sec:architecture}). Then,
$$\hat{w}=\displaystyle\argmin_{w\in\mathbb{R}^n}\sum_{u=1}^{N_{Train}}\sum_{i=1}^{n} \sum_{j=1}^m \left(F(\theta_u,w)_{ij}-F^*(\theta_u)_{ij}\right)^2.$$
\item Solve $$\hat{\theta}:=\displaystyle\argmin_{\theta\in\Theta}\sum_{i=1}^n \sum_{j=1}^m(\widetilde{F}(\theta)_{ij}-\sigma^{MKT}_{BS}(T_i,k_j))^2.$$
\end{enumerate}
\begin{remark}
Notice that $\hat{w}(\Delta)$ depends on $\Delta$ implicitly, consequently so does $\widetilde{F}(\theta)=F(\theta,\hat{w}(\Delta))$ (hence the name implicit learning). This setting is similar to that of image recognition and exploits the structure of the data to reduce the complexity of the Network (see Section \ref{sec:Numerics} for details). 
\end{remark}
\begin{remark}In our experiments we chose $n=8$ and $m=11$.
At first, a  criticism  of mapping \eqref{eq:implicitNetwork} might be the inability to extrapolate/interpolate between maturities/strikes outside the grid $\Delta$. However, one is free to choose the grids $\Delta$ as fine as needed. In addition, one may use standard (arbitrage free) uni/bi-variate splines techniques to extrapolate/interpolate across strikes and maturities, as with traditional market data observable only at discrete points.
\end{remark}
\begin{figure}[H]
\hspace*{0cm}
\caption{Volatility surface generated by the neural network approximator and the corresponding original counterpart on a grid given by $8$ maturities and $11$ strikes.}
\label{fig:NNGeneratedSmiles}
\end{figure}

\subsubsection{Some exotic payoffs}
\label{subsec: exotic payoff}
Our framework extends to a number of exotic products such as:  Digital barriers, no-touch (or double no-touch) barrier, cliquets or autocallables.\\

\noindent We present some numerical experiments in Section \ref{sec:Barrier}, to demonstrate the pricing of digital barrier options.  More precisely, in Section \ref{sec:Barrier} we consider down-and-in such as down-and-out digital barrier options, the main building blocks of many Autocallable products. For a barrier level $B< S_0$ and maturity $T$ the payoff is given by:

\begin{align}
P^{Down-and-In}(B,T)=\mathbb{E}\left[\ind_{\{\tau_B\leq T\}}\right]\label{eq:Down-and-in}\\
P^{Down-and-Out}(B,T)=\mathbb{E}\left[\ind_{\{\tau_B\geq T\}}\right]\label{eq:Down-and-out}
\end{align}
where $\tau_B=\displaystyle\inf_{t}\{S_t=B\}$. In this setting, we may easily generate a grid for barrier levels and maturities $\Delta^{Barrier}:=\{B_i,T_j\}_{i=1,\;j=1}^{n,\;\;\;\;m}$ that we can fit in the objective function specified in \eqref{eq:generalOF}

\subsection{Network architecture and training}

\label{sec:architecture}
Motivated by the above analysis, we choose to set up the calibration in the implicit two-step approach. This involves a separation of the calibration procedure into (i) ``Deep approximation" an approximation network with an implicit training and  (ii)``Calibration" a calibration layer on top. We first start by describing the approximation network in the implicit image-based training and discuss the calibration in Section \ref{sec:calibrationStep} below. In addition, we will highlight specific techniques that contribute to the robustness and efficiency of our design.

\subsubsection{Network architecture of the implied volatility map approximation}
Here we motivate our choice of network architecture for the following numerical experiments which were inspired by the analysis in the previous sections. Our network architecture is summarised in the graph \ref{im:Network} below.
\begin{enumerate}
\item A fully connected feed forward neural network with 4 hidden layers (due to Theorem \ref{thm:depth}) and $30$ nodes on each layers (see Figure \ref{im:Network} for a detailed representation)
\item Input dimension = $n$, number of  model parameters
\item Output dimension = 11 strikes$\times$ 8 maturities for this experiment, but this choice of grid can be enriched or modified.
\item The four inner layers have $30$ nodes each, which adding the corresponding biases results on a number $$(n+1)\times30+ 4\times (1+30)\times 30+(30+1)\times88=30 n+6478$$ of network parameters to calibrate (see Section \ref{sec:neuralnetworks} for details).
\item Motivated by Theorem \ref{thm:UniveralApproximationDerivatives} we choose the Elu $\sigma_{Elu}=\alpha(e^x-1)$ activation function for the network. 
\end{enumerate}

%
%
%

\begin{figure}[H]\hspace*{-1cm}\label{im:Network}
\begin{tikzpicture}[scale=0.8,every node/.style={scale=0.75},plain/.style={scale=0.8,
  draw=none,
  fill=none,
  },
  dot/.style={scale=086,draw,shape=circle,minimum size=3pt,inner sep=0,fill=black 
    },
net/.style={scale=0.8,
  matrix of nodes,
  nodes={
    draw,
    circle,
    inner sep=10pt
    },
  nodes in empty cells,
  column sep=1cm,
  row sep=-9pt
  },
>=latex
]
\matrix[net] (mat)
{
|[plain]| \parbox{1.3cm}{\large\centering Input\\layer} 
& |[plain]| \parbox{1.3cm}{\large\centering 1st Hidden\\layer} 
& |[plain]| \parbox{1.3cm}{\large\centering 2nd Hidden\\layer}
& |[plain]| \parbox{1.3cm}{\large\centering 3rd Hidden\\layer}  
& |[plain]| \parbox{1.3cm}{\large\centering 4th Hidden\\layer}  
& | [plain]| \parbox{1.3cm}{\large\centering Output\\layer} \\
|[plain]| &|[plain]| &|[plain]| &|[plain]| &|[plain]| & \\
|[plain]| &|[plain]| &|[plain]| &|[plain]| &|[plain]| & |[plain]|\\
|[plain]| & {$1$}  & {$1$} & {$1$} &{$1$} & \\
|[plain]|  \\
{$\theta_1$}& {\vdots} &{\vdots} & {\vdots} & {\vdots} &\\
  |[plain]|  \\
{\vdots}& {\vdots} & {\vdots} & {\vdots} & {\vdots} &\\
  |[plain]| \\
{$\theta_n$}& {\vdots} & {\vdots} & {\vdots} & {\vdots} &\\
  |[plain]|   \\
|[plain]|& {$30$} & {$30$}& {$30$} &{$30$} &\\
|[plain]| &|[plain]| &|[plain]| &|[plain]| &|[plain]| &|[plain]| &|[plain]| \\
|[plain]| &|[plain]| &|[plain]| &|[plain]| & |[plain]| & \\    };

\draw[<-] (mat-6-1) -- node[above] {Input 1} +(-2cm,0);
\draw[<-] (mat-8-1) -- node[above] {\vdots} +(-2cm,0);
\draw[<-] (mat-10-1) -- node[above] {Input $n$} +(-2cm,0);
\foreach \ai in {6,8,10}
{\foreach \aii in {4,6,8,10,12}
  \draw[->] (mat-\ai-1) -- (mat-\aii-2);
}
\foreach \ai in {4,6,8,10,12}
{\foreach \aii in {4,6,8,10,12}
  \draw[->] (mat-\ai-2) -- (mat-\aii-3);
}
\foreach \ai in {4,6,8,10,12}
{
\foreach \aii in {4,6,8,10,12}
  \draw[->] (mat-\ai-3) -- (mat-\aii-4);
}
\foreach \ai in {4,6,8,10,12}
{
\foreach \aii in {4,6,8,10,12}
  \draw[->] (mat-\ai-4) -- (mat-\aii-5);
}
\foreach \ai in {4,6,8,10,12}
{
\foreach \aii in {2,4,6,8,10,12,14}
  \draw[->] (mat-\ai-5) -- (mat-\aii-6);
}
\draw[->] (mat-2-6) -- node[above] {Output 1} +(2cm,0);
\draw[->] (mat-4-6) -- node[above] {\vdots} +(2cm,0);
\draw[->] (mat-6-6) -- node[above] {\vdots} +(2cm,0);
\draw[->] (mat-8-6) -- node[above] {\vdots} +(2cm,0);
\draw[->] (mat-10-6) -- node[above] {\vdots} +(2cm,0);
\draw[->] (mat-12-6) -- node[above] {\vdots} +(2cm,0);
\draw[->] (mat-14-6) -- node[above] {Output 88} +(2cm,0);

\end{tikzpicture}
\caption{Our neural network architecture with 4 hidden layers and 30 neurons on each hidden layer, with the model parameters of the respective model on the input layer and with the $8 \times 11$ implied volatility grid on the output layer.}
\end{figure}
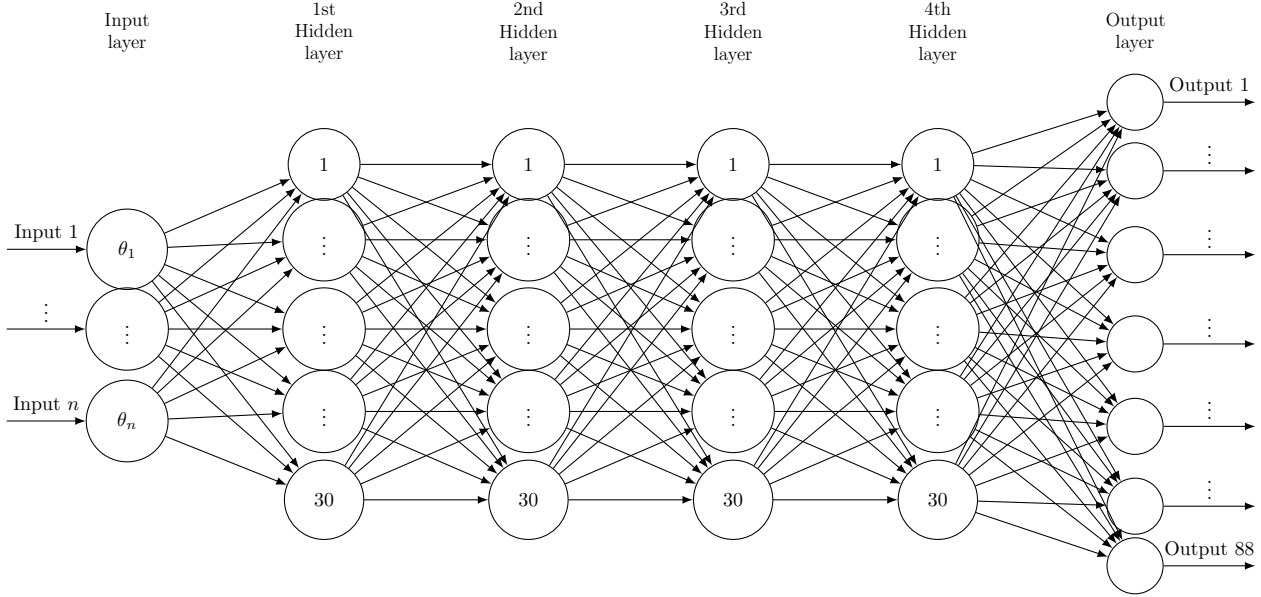

\subsubsection{Training of the approximation network}\label{sec:Training}

We follow the common features of optimization techniques and choose mini-batches, as described in Goodfellow, Bengio and Courville \cite{MLBook}.
Typical batch size values range from around 10 to 100. In our case we started with small batch sizes and increased the batch size until training performance consistently reached a plateau. 
Finally, we chose batch sizes of $32$, as performance is similar for batch sizes above this level, and larger batch sizes increase computation time  by computing a larger number of gradients at a time. 
\\

\noindent In our training design, we use a number of regularisation techniques to speed up convergence of the training, to avoid overfitting and improve the network performance.\\\\
\textbf{1) Early stopping:}
We choose the number of epochs as $200$ and stop updating network parameters if the error has not improved in the test set for $25$ steps.\\\\
\textbf{2) Normalisation of model parameters:}
Usually, model parameters are restricted to a given domain i.e. $\theta\in[\theta_{min},\theta_{max}]$. Then, we perform the following normalisation transform:
$$\frac{2\theta-(\theta_{max}+\theta_{min})}{\theta_{max}-\theta_{min}}\in[-1,1].$$

\textbf{3) Normalisation of implied volatilities:}
The normalisation of implied volatilities is a more delicate matter, since $\sigma_{BS}(T,k, \theta^{train})\in[0,\infty)$, for each $T$ and $k$. Therefore, we choose to normalise the surface subtracting the sample empirical mean and dividing by the sample standard deviation.

\subsection{The calibration step}
\label{sec:calibrationStep}
Once the pricing map approximation $\widetilde{F}$  for the implied volatility is found, only the calibration step in  \eqref{eq:ApproxCalibrationProblem} is left to solve.
In general, for financial models the  pricing map $F^*$ is assumed to be smooth (at least $C^1$ differentiable) with respect to all its input parameters $\theta$.\\
\vspace*{0.3cm}\\

\textbf{Gradient-based optimizers}\\\\
A standard necessary first order condition for optimality in \eqref{eq:ApproxCalibrationProblem} is that
\begin{equation}\label{eq:FOC}\nabla^{\theta} \delta\left(\widetilde{F}(\mathcal{M}(\theta),\zeta),\mathcal{P}^{MKT}(\zeta)\right)=0,\end{equation}
provided that the objective function is smooth. Then, a natural update rule is to move along the gradient via Gradient Descent i.e.

\begin{equation}\label{eq:GradientDescent}\theta_{i+1}=\theta_i-\lambda\nabla^{\theta} \delta\left(\widetilde{F}(\mathcal{M}(\theta_i),\zeta),\mathcal{P}^{MKT}(\zeta)\right),\quad \lambda>0.\end{equation} A common feature of gradient based optimization methods building on \eqref{eq:GradientDescent} is the use of the gradient $\nabla^{\theta} \delta\left(\widetilde{F}(\mathcal{M}(\theta),\zeta),\mathcal{P}^{MKT}(\zeta)\right)$, hence its correct and precise computation is crucial for subsequent success. Examples of such algorithms, are Levenberg-Marquardt \cite{Levenberg,Marquardt}, Broyden-Fletcher-Goldfarb-Shanno (BFGS) algorithm \cite{NumericalOptimization}, L-BFGS-B \cite{L-BFGS-B} and SLSQP \cite{SLSQP}. The main advantage of the aforementioned methods is the quick convergence towards condition \eqref{eq:FOC}. However, \eqref{eq:FOC} only gives necessary and not sufficient conditions for optimality, hence special care must be taken with non-convex problems.
\begin{remark}
Notably, making use of Theorem \ref{thm:UniveralApproximationDerivatives} we use a smooth activation functions in order to guarantee $\nabla^\theta \widetilde{P} \approx \nabla^\theta \widetilde{F}$\\
\end{remark}
\noindent\textbf{Gradient-free optimizers}\\\\
Gradient-free optimization algorithms are gaining popularity due to the increasing number of high dimensional nonlinear, non-differentiable and/or non-convex problems flourishing in many scientific fields such as biology, physics or engineering. As the name suggests, gradient-free algorithms make no $C^1$  assumption on the objective function. Perhaps, the most well known example is the Simplex based Nelder-Mead \cite{NelderMead} algorithm. However, there are many other methods such as  COBYLA \cite{COBYLA} or Differential Evolution \cite{DifferentialEvolution} and we refer the reader to \cite{GradientFreeReview} for an excellent review on gradient-free methods. The main advantage of these methods is the ability to find global solutions in \eqref{eq:ApproxCalibrationProblem} regardless of the objective function. In contrast, the main drawback is a higher computational cost compared to gradient methods.\\\\
To conclude, we summarise the advantages of each approach in Table \ref{table:gradientmethods}.
\begin{table}[H]
\centering
\begin{tabular}{c|c|c|}
\cline{2-3}
                                                                                                        & \textbf{Gradient-based}       & \textbf{Gradient-free} \\ \hline
\multicolumn{1}{|c|}{Convergence Speed}                       & Very Fast            & Slow          \\ \hline
\multicolumn{1}{|c|}{Global Solution}                                                                   & Depends on problem   & Always        \\ \hline
\multicolumn{1}{|c|}{\begin{tabular}[c]{@{}c@{}}Smooth activation \\ function needed\end{tabular}}      & Yes to apply Theorem \ref{thm:UniveralApproximationDerivatives} & No            \\ \hline
\multicolumn{1}{|c|}{\begin{tabular}[c]{@{}c@{}}Accurate gradient \\ approximation needed\end{tabular}} & Yes                  & No            \\ \hline
\end{tabular}
\caption{Comparison of Gradient vs. Gradient-free methods.}
\label{table:gradientmethods}
\end{table}

\section{Numerical experiments}\label{sec:Numerics}
In our numerical experiments we demonstrate that the accuracy of the approximation network indeed remains within the accuracy of the Monte Carlo error bounds and proclaimed in the introductory sections' objectives. For this we first compute the benchmark Monte Carlo errors in Figures \ref{picture:MCError1}-\ref{picture:MCError2} and compare this with the neural network approximation errors in Figures \ref{fig:rBergomi2} and \ref{fig:1FBergomi2}. For this separation into steps (i) and (ii) to be computationally meaningful, the neural network approximation has to be a reasonably accurate approximation of the true pricing functionals and each functional evaluation (i.e. evaluation an option price for a given price and maturity) should have a considerable speed-up in comparison to the original numerical method. In this section we demonstrate that our network achieves both of these goals. 
\subsection{Numerical accuracy and speed of the price approximation for vanillas}\label{subsec:arch}
\begin{figure}[H]
\hspace*{-1cm}
\includegraphics[scale=0.5]{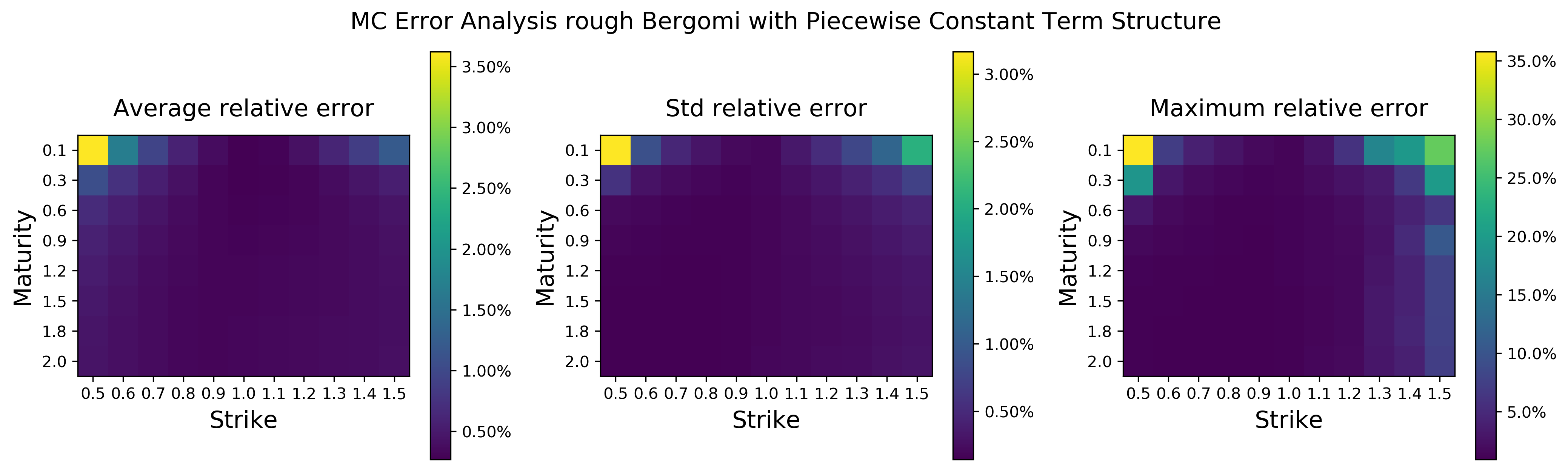}
\caption{As benchmark we recall average relative errors of Monte Carlo prices computed across $80,000$ random parameter combinations of the Rough Bergomi model. Relative errors are given in terms of Average-Standard Deviation-Maximum (Left-Middle-Right) on implied volatility surfaces in the Rough Bergomi model, computed using 95\% confidence intervals.}
\label{picture:MCError1}
\end{figure}

\begin{figure}[H] 
\hspace*{-1cm}\includegraphics[scale=0.5]{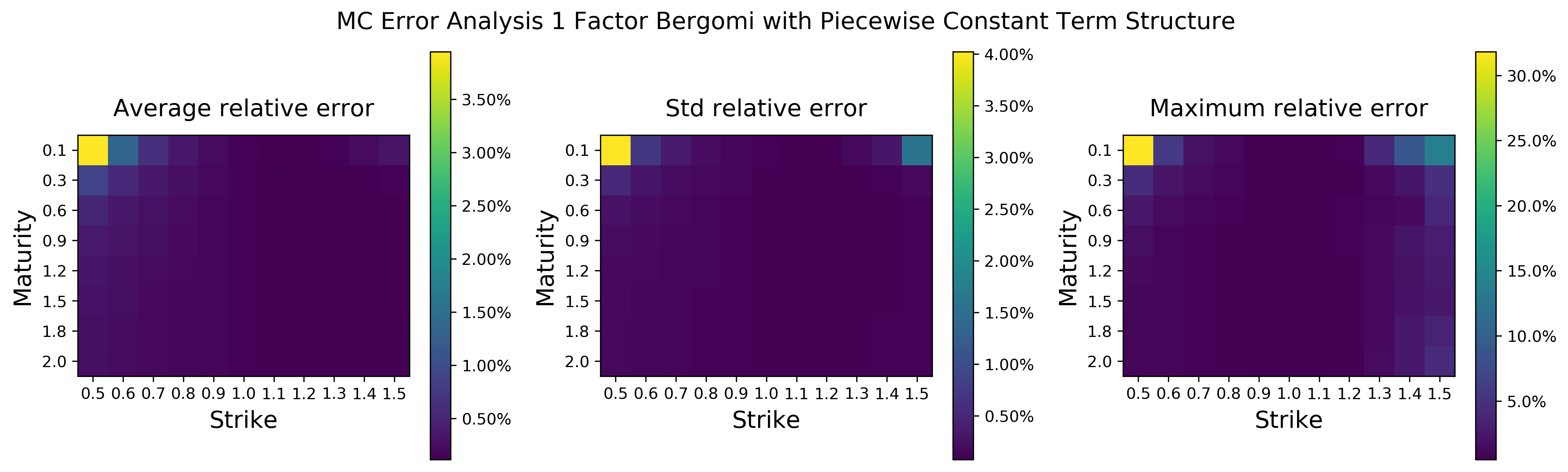}
\caption{As benchmark we recall average relative errors of Monte Carlo prices computed across $80,000$ random parameter combinations of the 1 Factor Bergomi model. Relative errors are given in terms of Average-Standard Deviation-Maximum (Left-Middle-Right) on implied volatility surfaces in the 1 Factor Bergomi model, computed using 95\% confidence intervals.}
\label{picture:MCError2}
\end{figure}

As mentioned in Section \ref{sec:Newperspectivecalib} one crucial difference that sets apart this work from direct neural network approaches, as pioneered by Hernandez \cite{Hernandez}, is the separation of (i) the implied volatility approximation function, mapping from parameters of the stochastic volatility model to the implied volatility surface--thereby bypassing the need for expensive Monte-Carlo simulations---and (ii) the calibration procedure, which (after this separation) becomes a simple deterministic optimisation problem.
As outlined in Section \ref{BypassingBottleneck} our aim for the Step (\textbf{i}) in the two-step training approach is to achieve a considerable speedup per functional evaluation of option prices while maintaining the numerical accuracy of the original pricer. Here we demonstrate how our NN training for Step (\textbf{i}) achieves these goals outlined in Section \ref{BypassingBottleneck}:
\begin{enumerate}
\item Approximation accuracy: here we compare the error of the approximation network error to the error of Monte Carlo evaluations. We compute Monte Carlo prices with $60,000$ paths as reference at the nodes where we compute the implied volatility grid using Algorithm 3.5 in Horvath, Jacquier and Muguruza \cite{HJM17}. 
In Figures \ref{picture:MCError1} and \ref{picture:MCError2} the approximation accuracy of the Monte Carlo method for the full implied volatility surface is computed using pointwise relative error with respect to the 95\% Monte Carlo confidence interval. Figures \ref{fig:rBergomi2} and \ref{fig:1FBergomi2} demonstrate that the same approximation accuracy for the neural network is achieved as for the Monte Carlo approximation (i.e. within a few basis points). For reference, the spread on options is around 0.2\% in implied volatility terms for the most liquid and those below a year. This translates into 1\% relative error for a implied volatility of 20\%.

\item  Approximation speed:
Table \ref{Table:NNSpeed} shows the CPU computation time per functional evaluation of a full surface under two different models; rBergomi \ref{example:rBergomi} and 1 Factor Bergomi \ref{example:1FBergomi} (for a reminder see Section \ref{sec:rBergomiGeneral} for details).  
\end{enumerate}
\begin{table}[H]
\hspace*{-1cm}
\centering
\begin{tabular}{c|c|c|c|c|c|}
\cline{2-6}
                                                                                                      & 
\multicolumn{1}{|c|}{\begin{tabular}[c]{@{}c@{}}MC Pricing\\1F Bergomi\\ Full Surface \end{tabular}}&
\multicolumn{1}{|c|}{\begin{tabular}[c]{@{}c@{}}MC Pricing\\rBergomi\\  Full Surface \end{tabular}} & \multicolumn{1}{|c|}{\begin{tabular}[c]{@{}c@{}}NN Pricing\\  Full Surface \end{tabular}}       & \multicolumn{1}{|c|}{\begin{tabular}[c]{@{}c@{}}NN Gradient\\ Full Surface \end{tabular}}   & \multicolumn{1}{|c|}{\begin{tabular}[c]{@{}c@{}}Speed up\\ NN vs. MC \end{tabular}}  \\ \hline
\multicolumn{1}{|c|}{\begin{tabular}[c]{@{}c@{}}Piecewise constant\\ forward variance\end{tabular}}  & $300,000\;\mu \text{s}$    & $500,000\;\mu \text{s}$ &$30.9\;\mu \text{s}$ & $113\;\mu \text{s}$ & $9,000-16 ,000$  \\ \hline
\end{tabular}
\caption{Computational time of pricing map (entire implied volatility surface) and gradients via Neural Network approximation and Monte Carlo (MC). If the forward variance curve is a constant value, then the speed-up is even more pronounced}
\label{Table:NNSpeed}
\end{table}

\subsubsection{Neural network price approximation in (rough) Bergomi models with piecewise constant forward variance curve}\label{sec:rBergomiGeneral}
We consider a piecewise constant forward variance curve  $\xi_0(t)=\sum_{i-1}^n \xi_i \ind_{\{t_{i-1}<t<t_i\}}$ where $t_0=0<t_1<...<t_n$ and $\{t_i\}_i=1,..,n$ are the option maturity dates ($n=8$ in our case). This is the modelling approach suggested by Bergomi \cite{BergomiBook}. We will consider again the rough Bergomi \ref{example:rBergomi} and 1 Factor Bergomi models \ref{example:1FBergomi}

\begin{itemize}
\item Normalized parameters as input  and normalised implied volatilities as output
\item 4  hidden layers with 30 neurons and \textit{Elu} activation function
\item Output layer with \textit{Linear} activation function
\item Total number of parameters: $6808$
\item Train Set: $68,000$ and Test Set: $12,000$
\item Rough Bergomi sample:$(\xi_0,\nu,\rho,H)\in \mathcal{U}[0.01,0.16]^8\times \mathcal{U}[0.5,4.0]\times \mathcal{U}[-0.95,-0.1]\times \mathcal{U}[0.025,0.5]$
\item 1 Factor Bergomi sample:$(\xi_0,\nu,\rho,\beta)\in \mathcal{U}[0.01,0.16]^8\times \mathcal{U}[0.5,4.0]\times \mathcal{U}[-0.95,-0.1]\times \mathcal{U}[0,10]$
\item strikes=$\{0.5,0.6,0.7,0.8,0.9,1,1.1,1.2,1.3,1.4,1.5\}$
\item maturities=$\{0.1,0.3,0.6,0.9,1.2,1.5,1.8,2.0 \}$
\item Training data samples of Input-Output pares are computed using Algorithm 3.5 in Horvath, Jacquier and Muguruza \cite{HJM17} with $60,000$ sample paths and the spot martingale condition i.e. $\mathbb{E}[S_t]=S_0,\;t\geq 0$ as control variate.
\end{itemize}
\begin{figure}[H]
\hspace*{-1cm}\includegraphics[scale=0.5]{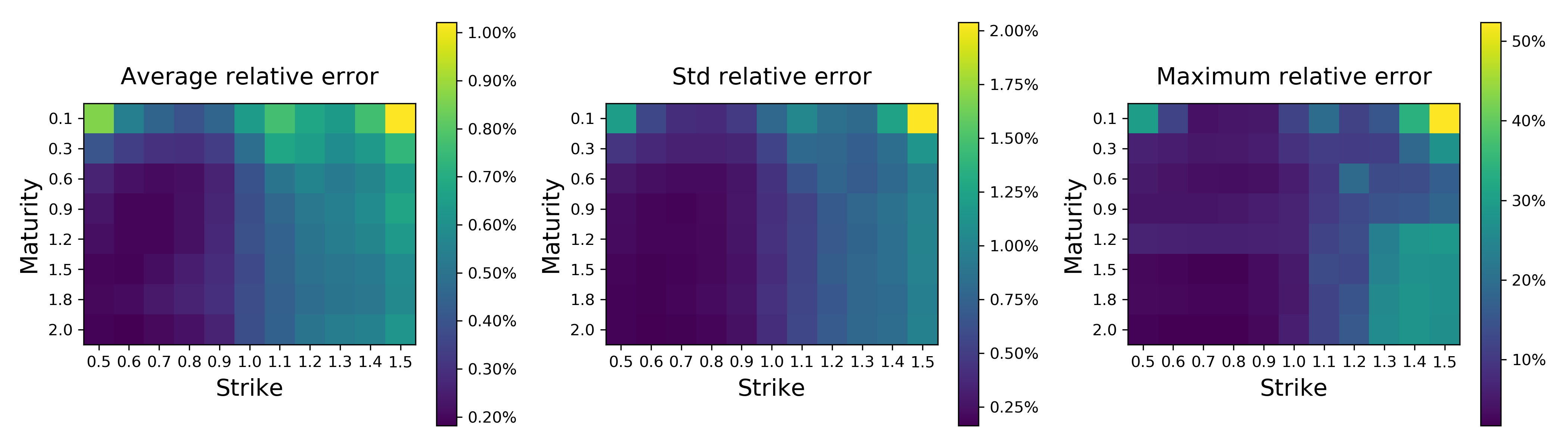}
\caption{We compare surface relative errors of the neural network approximator against the Monte Carlo benchmark across all training data ($68,000$ random parameter combinations)in the rough Bergomi model. Relative errors are given in terms of Average-Standard Deviation-Maximum (Left-Middle-Right).}
\label{fig:rBergomi2}
\end{figure}
\begin{figure}[H]
\hspace*{-1cm}\includegraphics[scale=0.5]{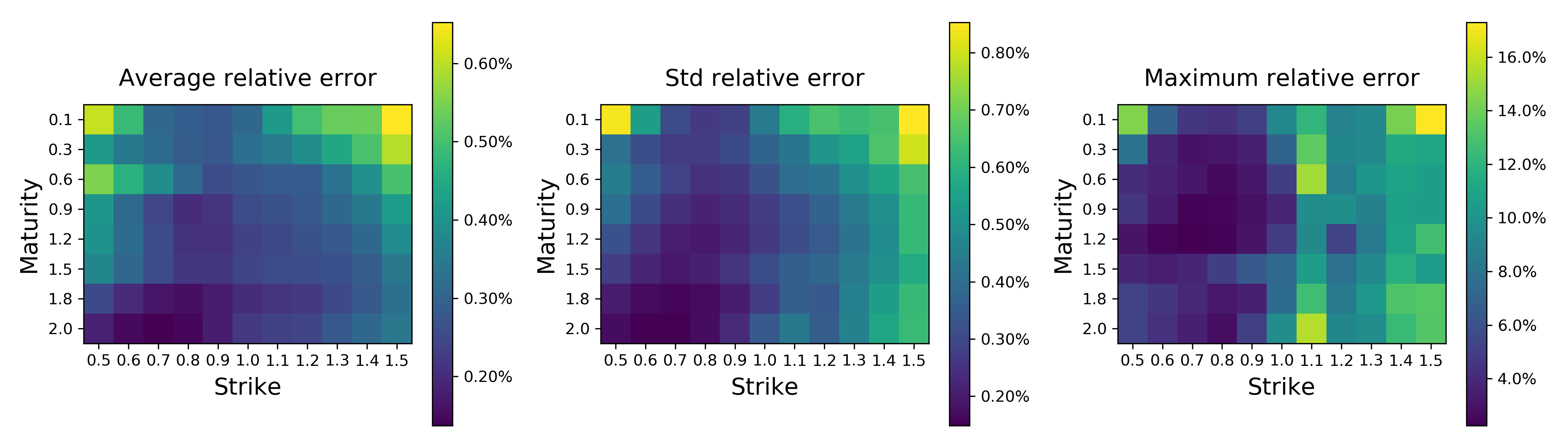}
\caption{We compare surface relative errors of the neural network approximator against the Monte Carlo benchmark across all training data ($68,000$ random parameter combinations)in the 1 Factor Bergomi model. Relative errors are given in terms of Average-Standard Deviation-Maximum (Left-Middle-Right).}
\label{fig:1FBergomi2}
\end{figure}

Figures  \ref{fig:rBergomi2} and \ref{fig:1FBergomi2} show that the average (across all parameter combinations) relative error between neural network and Monte Carlo approximations is far less than $0.5\%$ consistently (left image in Figures  \ref{fig:rBergomi2} and \ref{fig:1FBergomi2}) with a standard deviation of less than  $1\%$  (middle image in Figures  \ref{fig:rBergomi2} and \ref{fig:1FBergomi2}). The maximum relative error goes as far as $25\%$. We conclude that the methodology generalises adequately to the case of non-constant forward variances, by showing the same error behaviour.\\

\subsection{Calibration speed and accuracy for implied volatility surfaces}
\begin{figure}[H]
\hspace*{-2cm}\centering\includegraphics[scale=0.4]{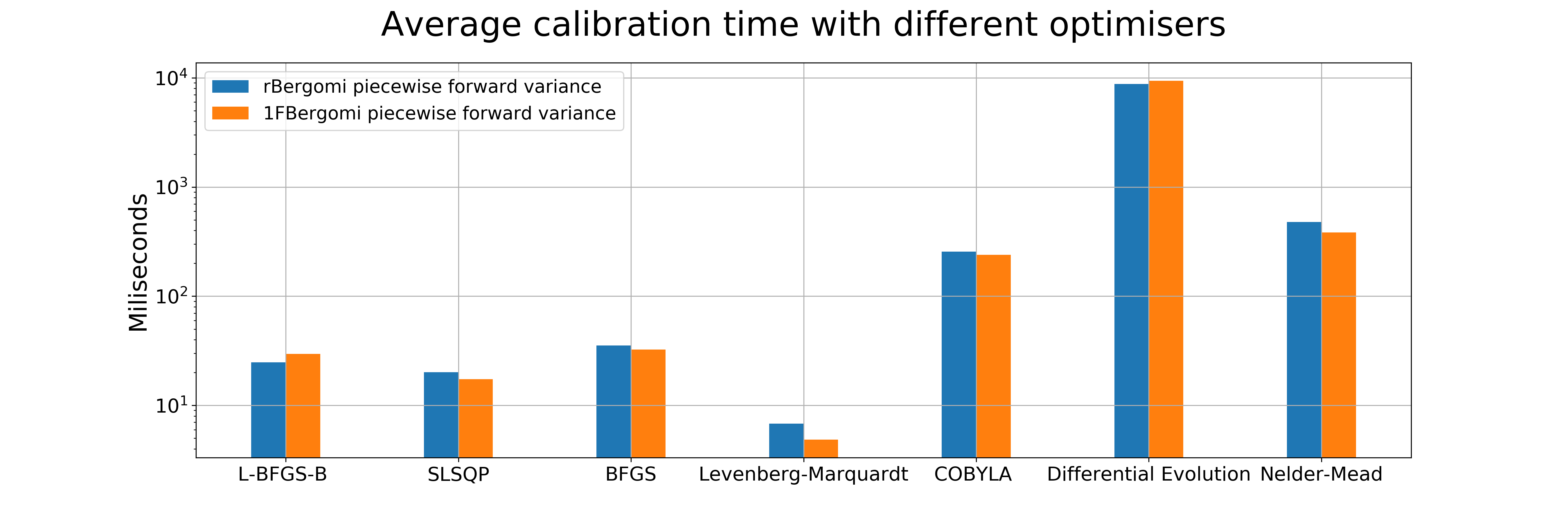}
\caption{Average calibrations times for all models using a range of optimizers.}
\label{fig:CalibrationTime2}
\end{figure}

Figure \ref{fig:CalibrationTime2} reports average calibration times on test set for different parameter combinations on each of the models analysed. We conclude that gradient-based optimizers outperform (in terms of speed) gradient-free ones. Moreover, in Figure  \ref{fig:CalibrationTime2} one observes that computational times in gradient-free methods are heavily affected by the dimension of the parameter space, i.e. flat forward variances are much quicker to calibrate than piecewise constant ones. We find that Lavenberg-Marquardt is the most balanced optimizer in terms of speed/convergence and we choose to perform further experiments with this optimizer. The reader is encouraged to keep in mind that a wide range of optimizers is available for the calibration and the optimal selection of one is left for future research.

In order to assess the accuracy, we report the calibrated model parameters $\widehat{\theta}$ compared to the synthetically generated data with the set of parameters $\overline{\theta}$ that was chosen for the generation of our synthetic data. We measure the accuracy of the calibration via parameter relative error i.e.
$$E_R(\widehat{\theta})=\frac{|\widehat{\theta}-\overline{\theta}|}{|\overline{\theta}|}$$ as well as the root mean square error (RMSE) with respect to the original surface i.e.
$$\text{RMSE}(\widehat{\theta})=\sqrt{\sum_{i=1}^n \sum_{j=1}^m(\widetilde{F}(\widehat{\theta})_{ij}-\sigma^{MKT}_{BS}(T_i,k_j))^2}.$$ Therefore, on one hand a measure of good calibration is a small RMSE. On the other hand, a measure of parameter sensitivity on a given model is the combined result of RMSE and parameter relative error.

\subsubsection{A calibration experiment with simulated data in (rough) Bergomi models with piecewise constant forward variances}\label{sec:rBergomiGeneralCalib}
\begin{figure}[H]
\hspace*{-1cm}\includegraphics[scale=0.4]{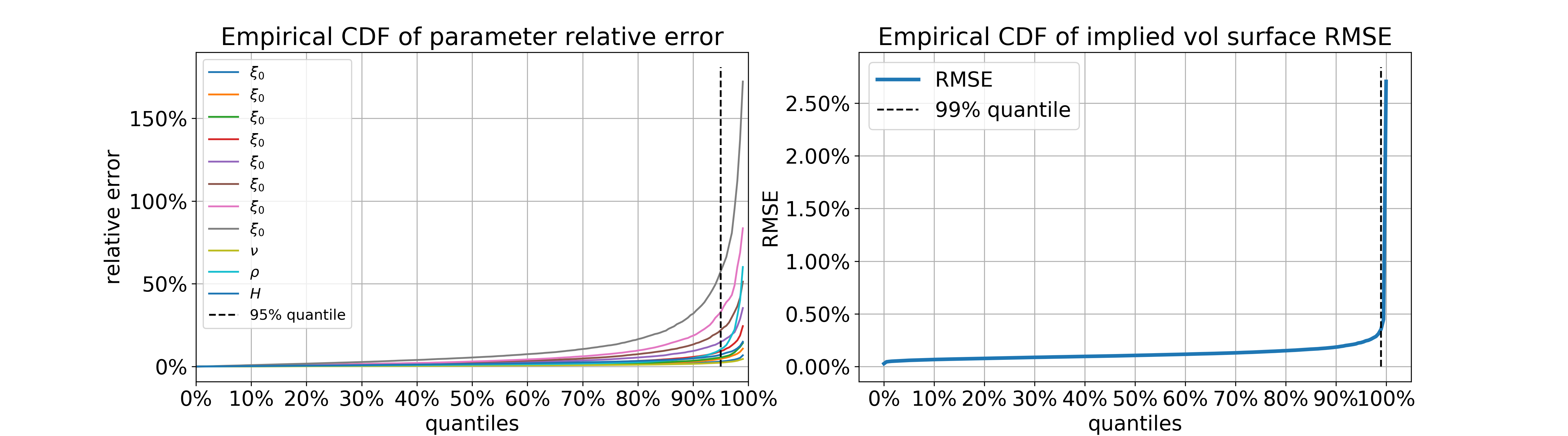}
\caption{Cumulative Distribution Function (CDF) of Rough Bergomi parameter relative errors (left) and RMSE (right) after Levengerg-Marquardt calibration across test set random parameter combinations.}
\label{fig:rBergomiTermStructure4}
\end{figure}

\begin{figure}[H]
\hspace*{-1cm}\includegraphics[scale=0.4]{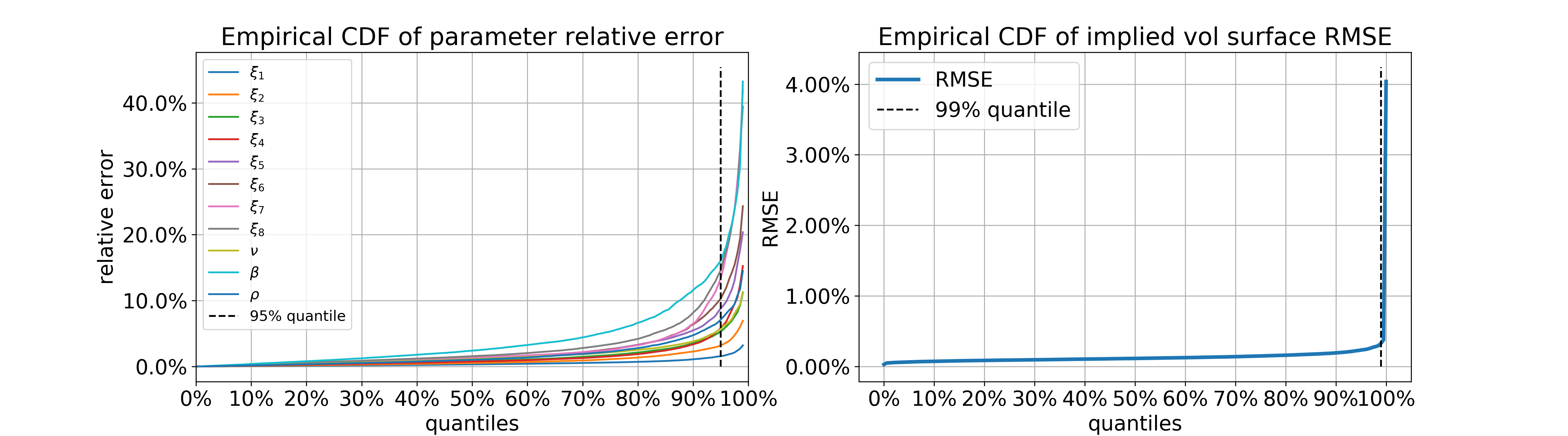}
\caption{Cumulative Distribution Function (CDF) of 1 Factor Bergomi parameter relative errors (left) and RMSE (right) after Levengerg-Marquardt calibration across test set random parameter combinations.}
\label{fig:1FBergomiTermStructure4}
\end{figure}

We consider the rough Bergomi model \eqref{example:rBergomi} and the Bergomi model \eqref{example:1FBergomi} with a piecewise constant term-structure of forward variances. Figures \ref{fig:rBergomiTermStructure4} and \ref{fig:1FBergomiTermStructure4} show that the $99\%$ quantile of the RMSE is below $1\%$ and shows that the Neural Network approach generalises properly to the piecewise constant forward variance. Again, we find that the largest relative errors per parameter are concentrated around $0$, consequence of using the relative error as measure. This suggests a successful generalisation to general forward variances, which to our knowledge has not been addressed before by means of neural networks or machine learning techniques.

\subsubsection{Calibration in the rough Bergomi model with historical data}
\label{sec:Historical}
\begin{figure}[H]
\centering
\includegraphics[scale=0.40]{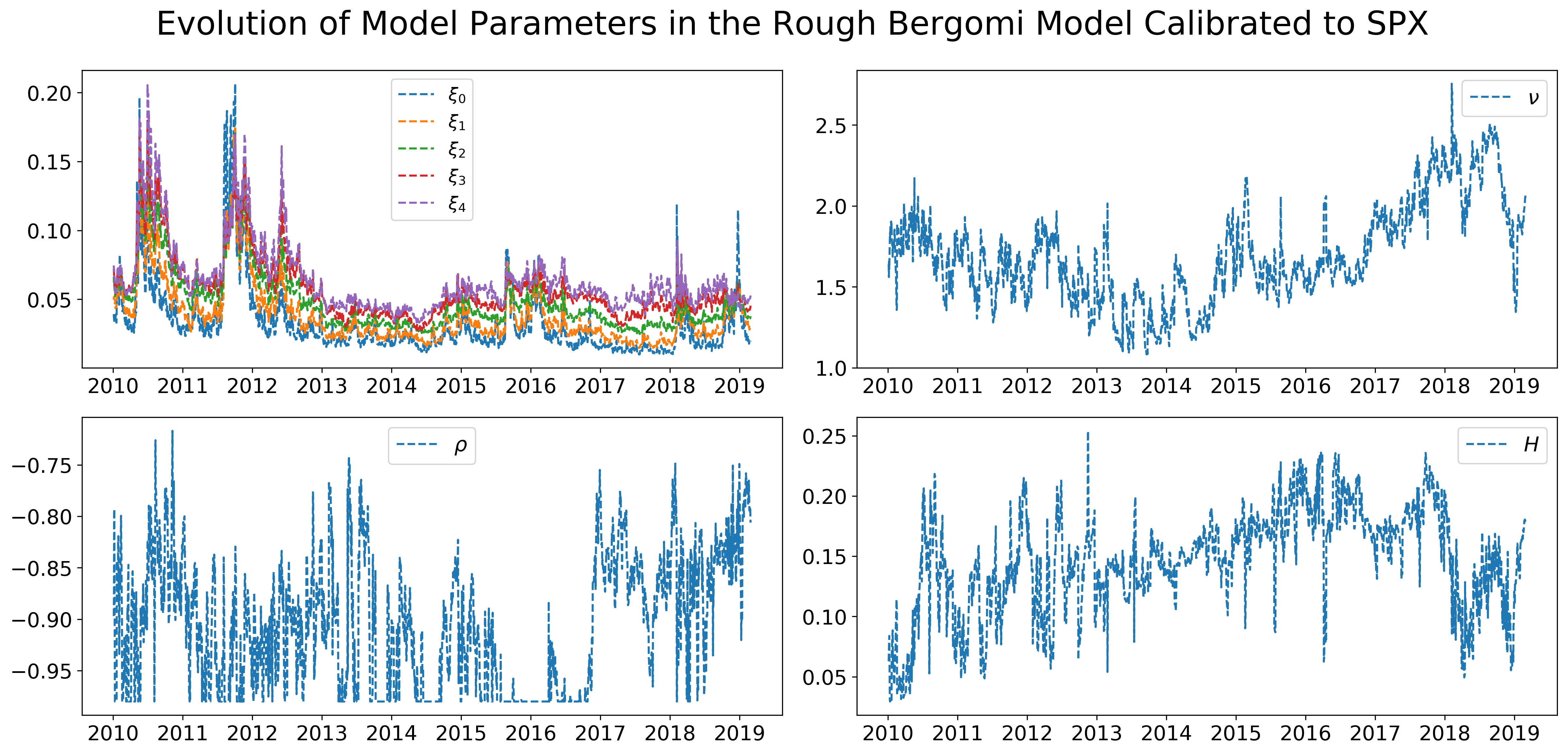}
\caption{Historical Evolution of parameters in the rough Bergomi model with a piecewise constant forward variance term structure calibrated on SPX}
\label{fig:HistoParams}
\vspace*{0.5cm}
\centering\includegraphics[scale=0.35]{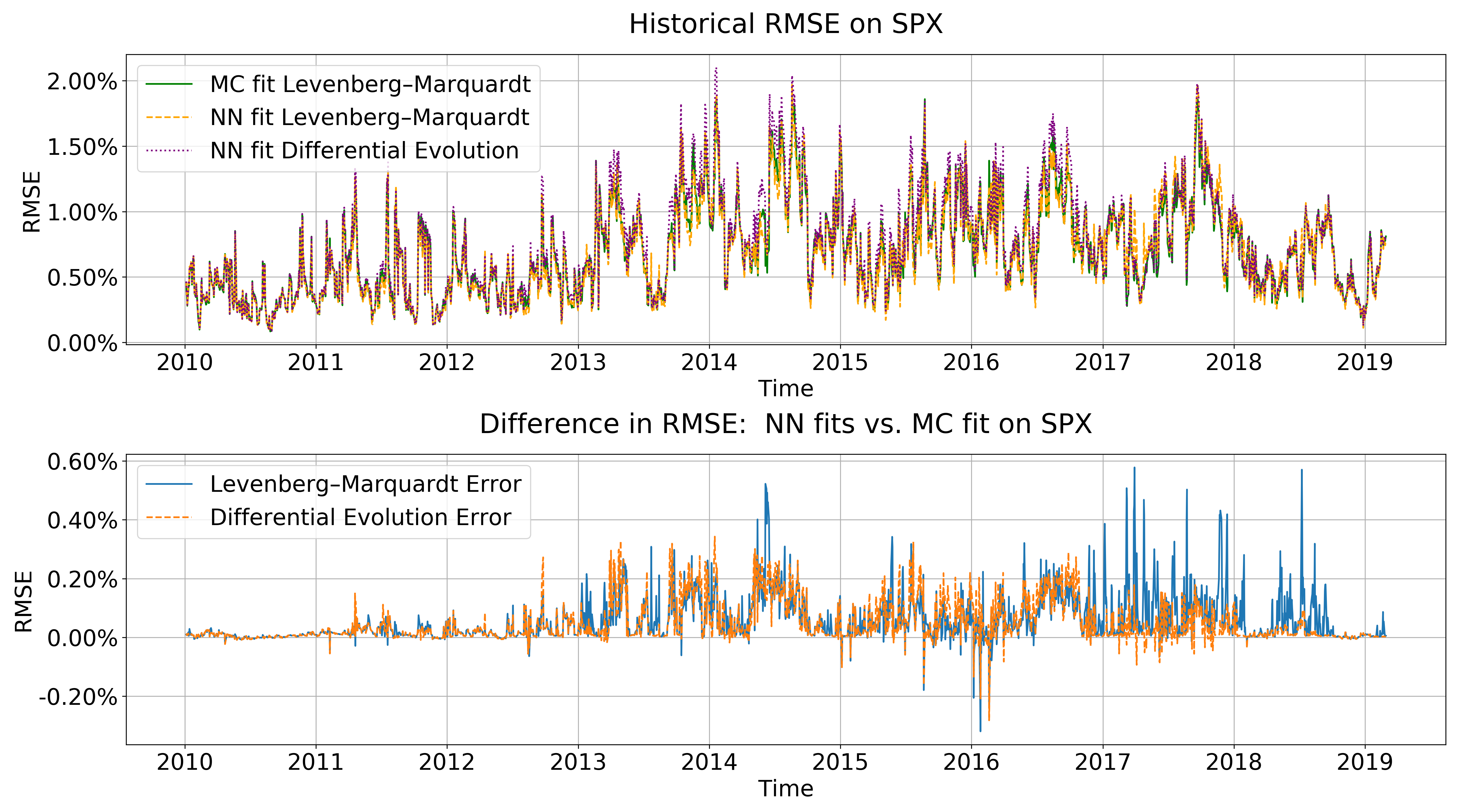}
\caption{The image above compares historical RMSE obtained by the neural network best  fit via Levenberg-Marquardt (dashed orange line) and Differential Evolution (dotted purple line) against the brute force MC calibration (green line) via Levenberg-Marquardt. Picture below shows the difference against MC brute force calibration.}
\label{fig:HistoRMSE3}
\end{figure}

As previously mentioned, the natural use of neural network approximators is the model calibration to historical data. We discussed that as along as the approximation is accurate, the calibration task should be performed within the given tolerance. Furthermore, one should expect such tolerance to be aligned with the neural network accuracy obtained in both training and test sets.\\\\
In this section we will perform a historical calibration using the neural network approximation and compare it with that of the brute force monte carlo calibration. Precisely we seek to solve the following optimisation problem for the rough Bergomi model
$$\hat{\theta^{rBergomi}}:=\displaystyle\argmin_{\theta^{rBergomi}\in\Theta^{rBergomi}}\sum_{i=1}^5 \sum_{j=1}^9(\widetilde{F}(\theta)_{ij}-\sigma^{MKT}_{BS}(T_i,k_j))^2.$$
where $\theta^{rBergomi}=(\xi_1,\xi_2,\xi_3,\xi_4,\xi_5,\nu,\rho,H)$ and $\Theta^{rBergomi}=[0.01,0.25]^5\times [0.5,4]\times [-1,0]\times [0.025,0.5]$. As for the time grid we choose $$(T_1,T_2,T_3,T_4,T_5):=\frac{1}{12}\times(1,3,6,9,12)$$ and for the strike grid $$k_i:=0.85+(i-1)\times 0.05 \quad \text { for } i=1,...,9.$$
We consider SPX market smiles between 01/01/2010 and 18/03/2019 on the pre-specified time and strike grid. Figure \ref{fig:HistoParams} shows the historical evolution of rough Bergomi parameters calibrated to SPX using the neural network price. In particular we note that $H<\frac{1}{2}$ as previously discussed in many academic papers \cite{AGM18, AlosLeon,BFG15,BFGMS17, BFGHS, BLP16, ER16, Fukasawa, GJR14, JMM17, HJL,  JPS17}, moreover we may confirm that under $\mathbb{Q}$, $H\in[0.1,0.15]$ as found in Gatheral, Jaisson and Rosenbaum \cite{GJR14} under $\mathbb{P}$. Figure \ref{fig:HistoRMSE3}, benchmarks the NN optimal fit using Levenberg-Marquardt and Differential Evolution against a brute force MC calibration via Levenberg-Marquardt. Again, we find that the discrepancy between both is below $0.2\%$ most of the time and conclude that the Differential Evolution algorithm does outperform the Levenberg-Marquardt. This in turn, suggests that the neural network might not be precise enough on first order derivatives. This observation, is left as an open question for further research. Perhaps surprisingly, we sometimes obtain a better fit using the neural network than the MC pricing itself. This could be caused by the fact that gradients in the neural network are exact, whereas when using MC brute force calibration we resort to finite differences to approximate gradients.

\subsection{Numerical experiments with barrier options in the rough Bergomi model}
\label{sec:Barrier}
\begin{figure}[H]
\hspace*{-1.5cm}\includegraphics[scale=0.46]{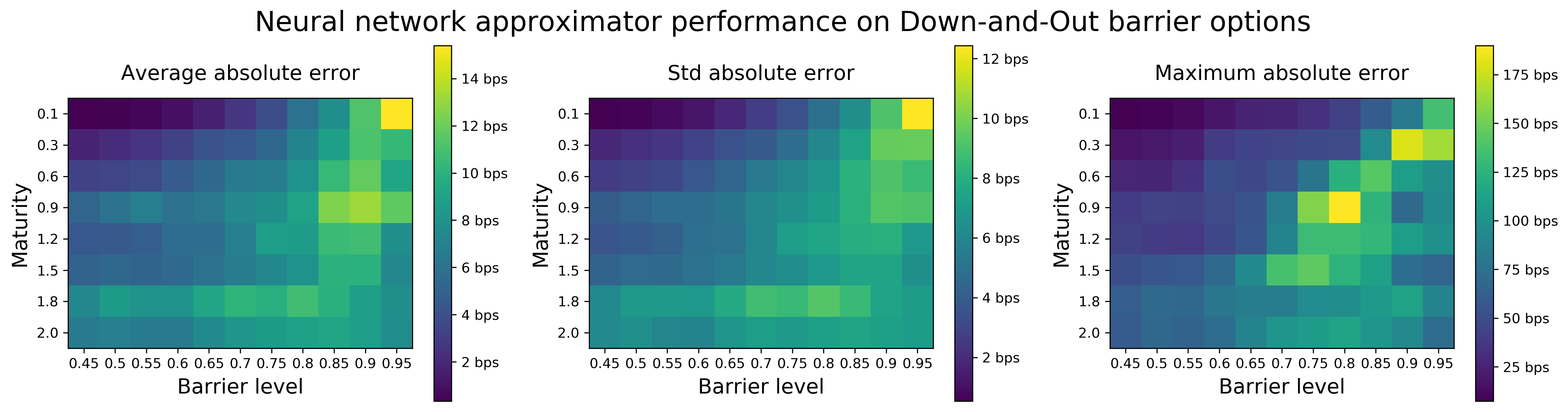}
\vspace*{0.5cm}
\hspace*{-1.5cm}\includegraphics[scale=0.46]
{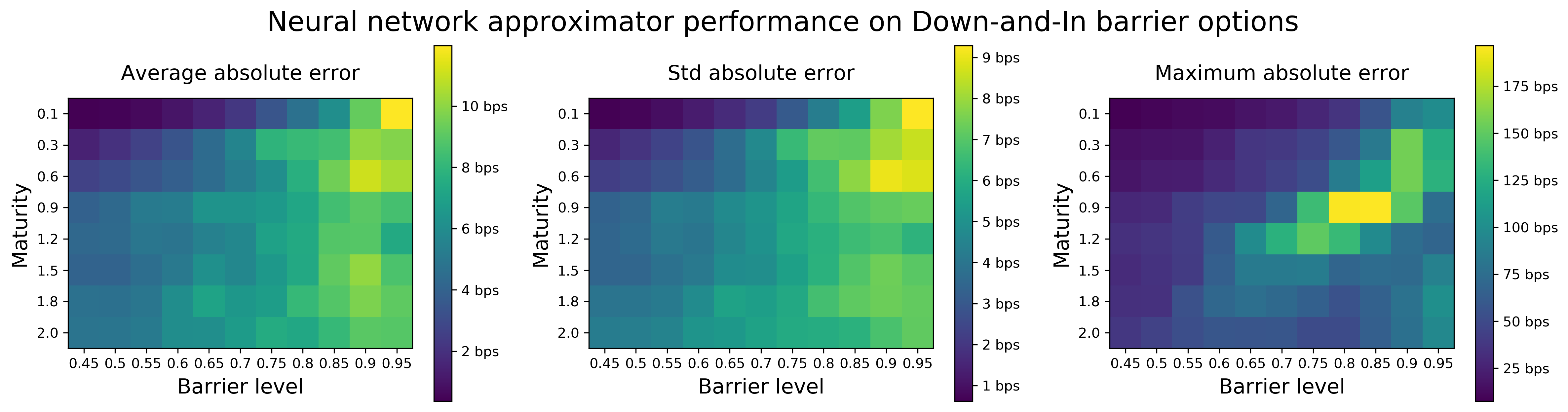}

\caption{Picute above: Down-and-Out neural network absolute error analysis on test set. Picture below: Down-and-In neural network absolute error analysis on test set }
\label{fig:Down-and-in}
\end{figure}

In this section we show that our methodology can be easily extended to exotic options. To do so we test our image-based approach on digital barrier options. We follow the same architecture and experimental design described in Section \ref{subsec:arch} for the rough Bergomi model. As described in Section \ref{subsec: exotic payoff} we adapt the objective function to the payoffs given in \eqref{eq:Down-and-in} and \eqref{eq:Down-and-out} and replace the strike grid by a barrier level grid. Figure \ref{fig:Down-and-in} confirm the accuracy of the neural network approximation with average absolute errors of less than 10bps with standard deviation of 10bps.

\section{Conclusions and outlook: ``best-fit'' models}\label{sec:Conclusions}
\noindent To sum up, neural networks have the potential to efficiently  approximate complex functions, which are difficult to represent and time-consuming to evaluate by other means. Using deep neural networks, as we will do here, to approximate the pricing map (or equivalently the implied volatility mapping) from  parameters of traditional models to shapes of the implied volatility surface represented by grid of implied volatility values speeds up each functional evaluation, while maintaining control over reliability and interpretability of network outputs. The implicit grid based approach that we advocate here, also allows further applications that opens up further landscapes for financial modelling.\\

\noindent \textbf{Potential applications and outlook towards mixture of ``expert" models}: In the previous sections we set up a powerful approximation method to closely approximate implied volatilities under different stochastic models and highlighted that the choice of the objective function (evaluation of the surface on a grid, inspired by pixels of an image) was crucial for the performance of the network. Now we are interested in the inverse task and ask whether a neural network---trained by this objective function to multiple stochastic models simultaneously---can identify which stochastic model a given set of data comes from. By doing so, potential applications we have in mind are twofold: \\
(1) Ultimately we are interested in which model (or what mixture of existing stochastic models) best describes the market.\\
(2) From a more academic and less practical perspective, we are interested whether and to what extent is it possible to``translate" parameters of one stochastic model to parameters of another.\\

\noindent We conduct a further, preliminary experiment as a proof of concept in the classification setting. We train a further neural network to identify which of three given stochastic volatility model generated a given implied volatility surface.\\

\textbf{Training procedure:} Implied volatility surfaces in this experiment were generated by the Heston, Bergomi and rough Bergomi models (see Section \ref{sec:stochastic models} for a reminder). For each volatility surface, a ``flag" was assigned corresponding to the model (eg: 1 for Heston, 2 for Bergomi and 3 for rough Bergomi). The training set thus consists of surfaces of the form: $(\Sigma^{\mathcal{M}(\theta)}_{BS}, I)$, where $\mathcal{M}$ is one of the three models $\mathcal{M}^{\textnormal{Heston}}$, $\mathcal{M}^{\textnormal{Bergomi}}$, $\mathcal{M}^{\textnormal{rBergomi}}$, $\theta$ an admissible combination of parameters for that model (thus in $\Theta^{\textnormal{Heston}}$, $\Theta^{\textnormal{Bergomi}}$ or $\Theta^{\textnormal{rBergomi}}$) and $I$ the flag identifying the model which generated the surface ($I=1$ if $\mathcal{M} = \mathcal{M}^{\textnormal{Heston}}$, $I=2$ if $\mathcal{M} = \mathcal{M}^{\textnormal{Bergomi}}$ and $I=3$ if $\mathcal{M} = \mathcal{M}^{\textnormal{rBergomi}}$). 
\\

\noindent We define a mixture of these surfaces as $\Sigma^{\mathcal{M}^{\textnormal{Mixture}((a,b,c))}} := a \Sigma^{\mathcal{M}^{\textnormal{Heston}}} + b \Sigma^{\mathcal{M}^{\textnormal{Bergomi}}} + c \Sigma^{\mathcal{M}^{\textnormal{rough Bergomi}}}$, where $a,b,c \geq 0$ and $a+b+c = 1$.  So far the training is suitable for recognition of a single model surface (either $a=0,b=0, c=1$, $a=0, b=1, c=0$ or $a=1, b=0,c=0$). To generalise this to mixtures, we randomly select surfaces (one from each model) and compute the mixture surface $\Sigma^{\mathcal{M}^{\textnormal{Mixture}((a,b,c))}} = a \Sigma^{\mathcal{M}^{\textnormal{Heston}}} + b \Sigma^{\mathcal{M}^{\textnormal{Bergomi}}} + c \Sigma^{\mathcal{M}^{\textnormal{rough Bergomi}}}$. The corresponding probabilities are $(a, b, c=1-a-b)$.\\ 

\textbf{Network Architecture:} The classifier is a small, fully connected feedforward network for the same reasons as those outlined in section \ref{sec:architecture}. The network is composed of 2 hidden layers (of 100 and 50 output nodes respectively) with exponentially linear activation functions and an output layer with a softmax activation function. Thus, the output of the network represents the probabilities of a given surface belonging to a particular model. We used stochastic gradient descent with 20 epochs to minimize cross-entropy (the cross-entropy of two discrete distributions $(p,q)$ with $K$ possible distinct values is $H(p,q) := - \sum_{1 \leq i \leq K} p_i \log q_i.$). \\

 \textbf{A numerical experiment on model recognition:} We report on one of many experiments here as a proof of concept: We test the method on mixtures of rough Bergomi and Heston surfaces (hence setting $b=0$ in the training). To vary the type of mixtures generated, we chose $a \in \{0, 0.1, \cdots, 0.9, 1\}$. For each $a$, the mixture surface is computed as the convex combination of a randomly chosen surface from the rough Bergomi and the Heston model, and repeated 20 times. The training set has 320,000 surfaces. To further test the robustness of the model, validation surfaces were generated using a finer grid of mixture parameters: $a \in \{0, 0.05, \cdots, 0.95, 1\}$. In total, the validation set is made up of 105,000 surfaces.

\begin{figure}[H]
	\label{fig:err_acc_rBergomiHeston}
	\centering
	\hspace*{-1cm}\includegraphics[scale=0.7]{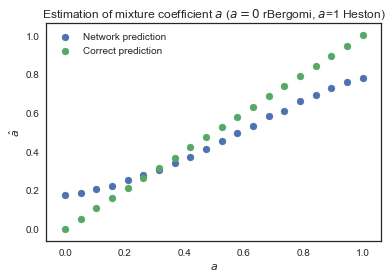}
	\caption{Error of the neural network classifier depending on the mixture coefficient $a$. Each point of the plot corresponds to the average estimated coefficient by the neural network for all mixture surfaces with a given $a$. For example, for $a=0$, the surfaces are generated from the rough Bergomi model. For each parameter combination from those surfaces, we compute the predicted mixture coefficient and average all of them over the validation set to report $\hat{a}$. The network never sets the mixture coefficient very close to $1$ or $0$, attributing the surface to one specific model. This may be explained using Bayesian reasoning.}
\end{figure}

\noindent We report the classifiers' effectiveness and comment on the results in Figure \ref{fig:err_acc_rBergomiHeston}.

\end{document}